# AI AND THE ECONOMY:
## AN ECONOMIC EXAMINATION OF
## PRODUCTION, DISTRIBUTION, FIRMS, LABOR, AND WELFARE

**Ali Zeytoon-Nejad, Ph.D.**
Associate Teaching Professor of Economics
Iglehart-Lightcap Faculty Fellow
School of Business
***Wake Forest University***
zeytoosa@wfu.edu

**Abstract:** Artificial Intelligence (AI) is rapidly transforming economic systems by altering production processes, labor markets, and the structure of firms and industries. AI is not merely a technological innovation. It is fundamentally a major economic phenomenon and a new wave of innovation with important implications for productivity, employment, market structure, public policy, long-run economic growth, and collective welfare. This essay examines the economics of AI by analyzing the multiple channels through which AI influences economic activity and societal well-being. It argues that AI should be understood simultaneously as a form of capital, a form of synthetic labor, a general-purpose technology, as well as an economic infrastructure. The analysis shows that AI functions as a multi-channel engine of productivity growth through automation, augmentation, optimization, prediction, and innovation. It is argued that AI has the potential to increase output, reduce costs, stimulate entrepreneurship, and accelerate economic growth, while also reshaping labor markets through labor substitution, labor complementarity, and creative destruction. The essay further examines AI's effects on competition, market concentration, and entrepreneurship. Finally, it explores the welfare implications of AI for consumers, producers, workers, governments, and society as a whole, and concludes by outlining policy considerations aimed at maximizing the benefits of AI while mitigating its potential adverse consequences.

**Keywords:** Artificial Intelligence; Economics of AI; Productivity; Economic Growth; Labor Markets; Creative Destruction; Industrial Organization; Welfare Economics

**JEL Classification:** O33, J24, D24, D43, L16

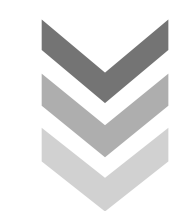

## 1. Introduction

Artificial Intelligence (AI) is emerging as one of the most transformative technologies in modern economic history. While technological progress has long served as a principal driver of productivity enhancement, economic growth, and improvements in living standards, AI possesses characteristics that distinguish it from many previous waves of innovation. Unlike traditional forms of technologies that primarily augment physical production processes, AI directly influences information processing, prediction, decision-making, creativity, and increasingly, autonomous action.[1] As a result, AI is not merely another technological advancement; it represents a fundamental shift in the way economic value is being created, distributed, and consumed.

Throughout history, major technological revolutions—including the Industrial Revolution, electrification, mass production, and the digital revolution—have reshaped economic systems by altering production technologies, labor demand, and industrial organization. Artificial Intelligence appears poised to exert a similarly transformative impact. However, unlike previous waves of automation that largely targeted '*manual and routine physical tasks*,' AI increasingly affects '*cognitive and analytical functions*' traditionally associated with highly educated workers. As a result, AI has the potential to influence virtually every sector of the economy, from manufacturing and agriculture to finance, healthcare, education, transportation, scientific research, and public administration, all of which involve cognitive and analytical functions.

[1]. AI may be understood as an integrated computational system built upon neural networks, deep learning, transformer architectures, large language models (LLMs), vision-language models (VLMs), reinforcement learning, massive datasets, high-performance computing (GPUs and TPUs), retrieval systems, and knowledge bases. Together, these technologies enable a broad collection of cognitive capabilities that were once provided by numerous separate software applications, including information retrieval, web searching and web scraping, writing assistance, word prediction, grammar and style correction, summarization, translation, coding, mathematical reasoning and computation, data analysis, planning, decision support, speech and voice recognition, visual recognition, video recognition, image and video generation, and many other forms of intelligent information processing.

The economic significance of AI extends beyond its capacity to automate existing tasks. From an economic standpoint, AI systems can generate predictions, optimize resource allocation, facilitate decision-making, discover patterns within massive datasets, and create entirely new products and services. These capabilities enable firms to reduce costs, improve efficiency, and grow output while simultaneously creating new business models and market opportunities. At the macroeconomic level, AI may contribute to higher productivity growth, accelerated innovation, and increased aggregate output. At the microeconomic level, AI alters the behavior of consumers, firms, and workers by changing information availability, transaction costs, and the economics of production.[2]

Despite its enormous potential, the rapid rise of AI also presents substantial economic changes and challenges. For instance, the emergence of AI is likely to change the distribution of production benefits across individuals, firms, industries, and nations. As another example, AI-driven automation may displace certain categories of labor while increasing demand for others. The resulting changes in labor market dynamics could alter wage structures, income distribution, and the relative shares of labor and capital in national income. Furthermore, because AI systems often exhibit strong economies of scale, network effects, and data-driven advantages, their widespread adoption may contribute to increased market concentration and the emergence of dominant firms with substantial economic and political influence due to the imbalance of power that they may benefit from.

These opportunities and challenges have stimulated growing interest in studying AI and its relationships with the economy. Economists, policy-makers, and business leaders increasingly seek to understand the dynamics of AI and how AI can affect markets, productivity, employment, wages, innovation, competition, economic growth,

[2]. Zeytoon-Nejad (2024) provides a comprehensive and detailed discussion of the economics of AI, and attends to the topic at the micro-, meso-, and macro-economic level.

and society's welfare overall. Yet many existing discussions remain fragmented, focusing on individual aspects such as labor displacement, technological unemployment, or productivity growth. A holistic economic framework is needed to integrate these various dimensions, evaluate the broader implications of AI for modern economies, and develop a big-picture understanding of all the dynamics being generated by AI in the economy.

This essay contributes to that objective by examining Artificial Intelligence through the lens of economic theory and economic-system analysis. The essay argues that AI should be viewed simultaneously as a form of capital, a general-purpose technology, and a new economic infrastructure for information processing and decision-making. These characteristics enable AI to influence both the supply side and the demand side of the economy while fundamentally altering the relationship between labor, capital, knowledge, and production.

More specifically, the essay addresses several key questions, examples of which include: How does AI affect production processes and economic efficiency? What impact may AI have on labor markets, employment patterns, and income distribution? How does AI influence market structure, competition, and industrial organization? What are the implications of AI for economic growth and social welfare? Finally, what policy frameworks can help maximize the benefits of AI in the long-run while mitigating its potential risks and adverse effects in the short-run?

A central argument advanced in this essay is that Artificial Intelligence has the potential to become one of the most important drivers of economic growth and productivity in the twenty-first century. However, its ultimate economic impact will depend not only on technological capabilities but also on the institutional, legal, and policy environments within which it is allowed to be developed and deployed. AI possesses the capacity to generate unprecedented levels of efficiency and prosperity in the long run, yet it may also create significant challenges in the short run and medium run related to labor adjustment, income distribution changes, market concentration, and governance.

The economic question is therefore not whether AI will transform the economy, but rather how its process of creative destruction will unfold over the short run and long run; who will emerge as its winners and losers; and how the gains and losses generated by its process of creative destruction will be distributed across individuals, firms, industries, and societies.

The remainder of the essay proceeds as follows. Section 2 develops the conceptual foundations of the economics of Artificial Intelligence and examines AI as a form of capital, synthetic labor, general-purpose technology, and economic infrastructure. Section 3 analyzes the effects of AI on innovation, productivity, and long-run economic growth. Section 4 investigates the implications of AI for labor markets, employment, and the future of work through the lenses of labor substitution, labor complementarity, and creative destruction. Section 5 examines the impact of AI on market structure, competition, entrepreneurship, and industrial organization. Section 6 evaluates the welfare implications of AI for consumers, producers, workers, governments, and society as a whole. Section 7 explores the broader implications of AI for the future evolution of economic systems, institutions, and human prosperity. Section 8 discusses policy implications and presents recommendations for maximizing the benefits of AI while mitigating its potential costs. Finally, Section 9 concludes and discusses the broader economic significance of Artificial Intelligence.

## 2. Conceptual Foundations: Artificial Intelligence as the New Engine of Economic Transformation

**(*What exactly is Artificial Intelligence from an economic perspective?)***

### 2.1 Understanding AI Beyond Technology

Artificial Intelligence is frequently described as a technological innovation. While this characterization is accurate, it is incomplete from an economic perspective. AI is not merely a new form of technology; it is a transformative economic phenomenon that is reshaping many economic processes and institutions, including production, consumption, innovation, labor markets, industrial organization, and the institutional mechanisms through which economic decisions are made.

Throughout economic history, transformative innovations have altered the structure of economic systems by changing the relationship between labor, capital, knowledge, and production. The steam engine expanded humanity's productive capacity by augmenting physical power. Electricity revolutionized industrial production and enabled entirely new industries. Computers dramatically reduced the cost of information processing, while the Internet transformed communication, commerce, and information exchange. Artificial Intelligence represents the next stage in this historical progression by enabling machines to perform tasks that previously required human cognition, including learning, prediction, problem solving, language processing, pattern recognition, and increasingly, creative and generative activities.

As a result, AI affects not only the quantity of output that can be produced but also the manner in which economic decisions are made, resources are allocated, and knowledge is created and utilized throughout society. As such, its influence extends across virtually every sector of the economy and reaches far beyond the boundaries of individual firms or industries.

To understand these broader implications, this essay proposes a multidimensional framework that views AI simultaneously as (1) a form of capital, (2) a form of labor, (3) an economic infrastructure, (4) a general-purpose technology, and (5) an emerging economic institution.

### 2.2 AI as the Latest Wave of Creative Destruction

Before examining the five mentioned dimensions individually, it is useful to place AI within the broader historical process of economic evolution. Schumpeter (1942) famously described capitalism as a process of "creative destruction" through which new technologies, firms, products, and organizational forms continually replace older ones. Economic progress does not emerge from stability alone; rather, it arises through continual experimentation, innovation, adaptation, and renewal.

Recent work by Zeytoon-Nejad (2026a) argues that market economies are characterized by two fundamental forms of dynamism: *business cycles* and *creative destruction*. While business cycles generate short-run fluctuations around a growth path, creative destruction alters the growth path itself through innovation, technological advancement, entrepreneurial discovery, and industrial transformation. In this framework, capitalism is best understood as an evolutionary system that continually reinvents itself through successive waves of innovation. Figure 1 represents these successive waves of innovation.

**Figure 1:** Creative Destruction as Six Waves of Innovation

**Note:** *Adapted from Zeytoon-Nejad (2026a). The figure illustrates the successive waves of innovation through which creative destruction drives long-run economic progress.*

Artificial Intelligence represents the most recent major wave in this process of creative destruction. Just as mechanization transformed agriculture, electrification transformed industry, automation transformed manufacturing, and digitalization transformed information exchange, AI is transforming cognition itself. It is reshaping how information is processed, how decisions are made, how products are designed, how services are delivered, and how knowledge is generated.

Consequently, AI should not be viewed merely as another productivity-enhancing technology. Instead, it should be understood as a major transformative force capable of restructuring industries, reallocating labor and capital, creating entirely new markets, and accelerating the evolutionary dynamics of the market economy. In this sense, AI is both a product of creative destruction and a powerful catalyst for future waves of creative destruction.

### 2.3 AI as Capital

The most direct economic interpretation of AI is as a form of capital, which is a factor of production. Like machinery, equipment, software, and other productive tools and capital goods, AI is created through investment over time and subsequently employed in the process of production to generate economic value. However, AI differs from traditional forms of capital in several important respects.

First, AI systems generally require substantial fixed costs of development but relatively low marginal costs of replication. Once an AI model has been trained, it can often be deployed across millions of users at minimal additional cost.

Second, AI exhibits extraordinary scalability. Unlike conventional capital assets that are often constrained by physical capacity, AI systems can frequently serve large numbers of users simultaneously.

Third, AI improves through learning and feedback. Additional data, usage, and refinement often increase the productive capabilities of AI systems over time. These characteristics create the possibility of increasing returns to scale and may significantly strengthen the competitive advantages of firms that successfully develop or acquire superior AI capabilities.

### 2.4 AI as Labor

Although AI may be owned and accumulated like capital, it increasingly performs tasks traditionally associated with labor, which is another factor of production. Historically, machines primarily ‘substituted’ for “physical effort” while ‘complementing’ “human cognitive abilities”. Artificial Intelligence differs because it increasingly ‘substitutes’ for cognitive work itself.

Activities such as data analysis, customer service, language translation, software development, document review, forecasting, pattern recognition, and content generation can now be performed partially or entirely by AI systems. From an economic perspective, AI may therefore be viewed as a form of synthetic labor, which is capable of doing a wide range of human cognitive tasks.

AI labor possesses characteristics that are distinctive from human workers. It can operate continuously without fatigue, be replicated almost instantaneously, perform tasks simultaneously across multiple locations, and scale rapidly with relatively low marginal costs.

The emergence of synthetic labor represents one of the most important economic developments of the twenty-first century because it extends automation into domains previously believed to be protected from technological substitution. Consequently, understanding the interaction between AI and human labor is central to understanding the future evolution of employment, labor markets, wages, and income distribution, which are to be discussed in greater detail in Section 4 and Appendix A.

### 2.5 AI as Human Capital

Beyond its role as capital and synthetic labor, Artificial Intelligence can also be viewed as a form of human capital. Human capital traditionally refers to the knowledge, skills, competencies, and expertise that individuals acquire through education, training, and experience. These capabilities enhance worker productivity and increase an individual's economic value in the labor market.[3]

Artificial Intelligence increasingly functions as a complement to human capital by expanding the productive capabilities of individuals who possess the knowledge and skills required to utilize it effectively. Just as literacy, numeracy, computer literacy, and digital skills became important forms of human capital during earlier technological eras, AI literacy and AI proficiency are emerging as increasingly valuable forms of human capital in the modern economy.[4]

Workers who understand how to leverage AI tools effectively can often perform tasks more quickly, more accurately, and at a larger scale than workers who lack such capabilities. AI-assisted professionals can analyze larger datasets, generate more sophisticated insights, automate routine tasks, accelerate research, improve decision-making, and increase overall productivity. Therefore, the economic value of a worker may increasingly depend not only on traditional education and experience but also on the ability to work effectively and productively with AI systems.

More importantly, AI may create new forms of human capital that are difficult to replicate. Individuals who develop specialized expertise in applying AI within particular domains may acquire unique capabilities that distinguish them from other workers. Such individuals may be able to produce outputs, solve problems, or create value in ways that others cannot easily match. As a result, they may obtain significant competitive advantages in labor markets, entrepreneurship, and professional careers.[5]

### 2.6 AI as Infrastructure

Artificial Intelligence can also be viewed as a form of economic infrastructure. Infrastructure consists of foundational systems that support broader economic activity. Transportation networks, electricity grids, telecommunications systems, and the Internet all serve this function.

Increasingly, AI is becoming embedded within the underlying operations of economic systems. AI supports logistics management, financial transactions, healthcare diagnostics, educational platforms, manufacturing processes, supply chains, scientific research, and government administration. As AI becomes integrated into these foundational systems, it increasingly resembles a general

[3]. While the previous section examined AI as a form of synthetic labor capable of performing productive tasks directly, this section considers AI from a different perspective: as a form of human capital. Here, the focus is not on AI replacing or complementing workers, but on how the ability to use AI effectively becomes a valuable skill that enhances worker productivity and labor-market competitiveness.

[4]. For readers interested in the origins and evolution of the concept of human capital in economics, the earliest conceptual discussion appears in Smith's analysis of the productive value of acquired skills and education (Adam Smith, 1776). The modern human capital literature was subsequently developed by Schultz (1961), who emphasized investment in education, health, and training, and by Becker (1964), who formalized human capital theory within microeconomic analysis. Within the economic growth literature, human capital was first incorporated into a formal growth model by Uzawa (1965). The endogenous growth literature was subsequently advanced by Romer (1986, 1990), whose pioneering work on endogenous technological change highlighted the complementary roles of knowledge, innovation, and human capital in sustaining economic growth.

[5]. This observation highlights an important distinction. The workers most vulnerable to AI may not necessarily be those competing directly against AI systems. Rather, they may be those competing against other workers who have successfully learned how to utilize AI. In many occupations, the relevant competition may increasingly be between AI-augmented workers and non-AI-augmented workers rather than between human workers and AI machines themselves. From this perspective, AI should be viewed not only as a technology that transforms production by replacing labor with capital but also as a new form of human capital that enhances the productivity, adaptability, and market value of those who acquire the knowledge and skills necessary to use it effectively. As AI continues to diffuse throughout the economy, AI-related competencies may become increasingly important determinants of individual productivity, earnings, and labor-market success.

infrastructure that facilitates activity throughout the economy rather than a standalone technology confined to specific firms or sectors.

The historical experience of electricity provides a useful analogy. The greatest economic benefits of electrification did not arise solely within the electricity industry itself; rather, they emerged through productivity improvements across virtually every industry. AI can follow a similar trajectory, generating value primarily through its widespread integration into the broader economy.

### 2.7 AI as a General-Purpose Technology

Economists often classify transformative innovations as General-Purpose Technologies (GPTs – not to be confused with Generative Pre-Trained Transformers which are a type of AI). Examples of GPTs include the steam engine, electricity, the internal combustion engine, computers, and the Internet.[6]

General-purpose technologies typically possess three defining characteristics:

1. Broad applicability across industries;
2. Continuous improvement over time; and
3. Strong complementarities with other innovations.

Artificial Intelligence satisfies all the three conditions. Its applications extend across nearly every economic sector. Its capabilities continue to improve as computational power, algorithms, and data availability expand. Moreover, AI enables complementary innovations in medicine, finance, transportation, manufacturing, education, scientific research, and countless other fields.

The significance of GPTs extends beyond their direct contributions to productivity. Their greatest economic impact often arises from the secondary innovations they enable. Thus, the long-run and indirect economic effects of AI may prove substantially larger than its immediate productivity gains.

### 2.8 AI as an Emerging Economic Institution

Perhaps the most novel perspective is to view AI as an emerging economic institution. Institutions are the rules, procedures, and decision-making structures that organize economic interactions. Markets, firms, legal systems, contracts, and governments all serve institutional functions.

Increasingly, AI systems participate directly in economic decision-making processes. AI algorithms influence risk assessments, credit allocation, insurance pricing, hiring decisions, goods visibility, product recommendations, consumer choice, financial transactions, and content moderation. In these contexts, AI is no longer merely a tool employed by institutions. Rather, it becomes part of the institutional mechanism itself.

As AI systems assume greater responsibility for making decisions, allocating resources, and shaping economic outcomes, they increasingly function as decentralized governance mechanisms embedded within markets and organizations. This development of an intermediary tool raises important questions regarding accountability, reliability, agency, transparency, fairness, and legitimacy that extend far beyond traditional concerns about the quality of decentralized or centralized economic outcomes.

### 2.9 A Unified Framework for the Economics of AI

One of the central arguments of this essay is that AI occupies a unique position within modern economic systems because it simultaneously embodies multiple economic roles. It is:

[6]. AI as infrastructure and AI as a general-purpose technology capture different aspects of the same phenomenon. The infrastructure perspective emphasizes AI as a ‘foundational platform’ that supports economic activity, while the general-purpose technology perspective emphasizes AI’s ability to transform production processes, generate complementary innovations, and ‘diffuse across a wide range of industries’. Thus, the former focuses on AI’s ‘foundational’ enabling role, whereas the latter focuses on its ‘widespread’ transformative role.

1. Capital that expands productive capacity;
2. Synthetic labor that performs cognitive tasks;
3. Human capital that creates a competitive advantage in labor markets for individuals with special AI- related knowledge;
4. Infrastructure that supports economic activity;
5. A general-purpose technology that enables widespread innovation; and
6. An emerging institution that influences different aspects of economic decision-making.

At the same time, AI represents the newest and perhaps most consequential wave of Schumpeterian creative destruction in the modern economy. It is accelerating the evolutionary dynamics through which market economies adapt, innovate, and transform themselves over time. Just as previous waves of innovation reshaped agriculture, manufacturing, transportation, and communication, AI is reshaping the economics of knowledge, cognition, and decision-making.

Understanding AI through this broader framework provides the conceptual foundation for the remainder of the essay. The sections that follow examine how AI influences productivity, innovation, labor markets, industrial organization, economic growth, and social welfare, and why its long-run impact may rival or exceed that of any previous technological revolution.

## 3. Artificial Intelligence, Productivity, and Economic Growth

***(How does Artificial Intelligence affect economic growth?)***

### 3.1 Introduction

Throughout economic history, sustained improvements in living standards have been driven primarily by increases in productivity. As proven by the findings of growth-accounting studies, while nations may also increase output through greater utilization of labor, capital, or natural resources, long-run, sustained economic growth primarily depends upon the ability to produce more value with the same—or fewer—inputs, a phenomenon known as productivity growth. Consequently, productivity growth occupies a central position in the theory of economic growth and development.

Artificial Intelligence has emerged as one of the most promising sources of productivity growth in the modern economy. By automating cognitive tasks, improving decision-making, accelerating innovation, and reducing information-processing costs, AI has the potential to increase efficiency across virtually every sector of the economy. Unlike many previous technologies that primarily enhanced physical production processes, AI directly affects the creation, processing, interpretation, and application of knowledge itself. As a result, its influence may extend to nearly every dimension of economic performance.

This section examines the mechanisms through which AI contributes to productivity growth and explores its potential implications for long-run economic growth and development.

### 3.2 AI and the Economics of Productivity

Let's start with an aggregate production function of the following form for an economy:

$$Y = TFP \cdot f(L, K, \dots)$$

where Y denotes quantity of production (i.e., economic output), which is dependent on the number of workers or simply labor (denoted by L), quantity of capital (denoted by K), and the level of Total Factor Productivity (TFP) of all the factors of production that are utilized in the process of production.

Figure 2 illustrates two distinct mechanisms through which Artificial Intelligence may increase aggregate output. The lower curve represents the economy's initial aggregate production function, while the upper curve represents the aggregate production function after AI has generated improvements in the TFP level.

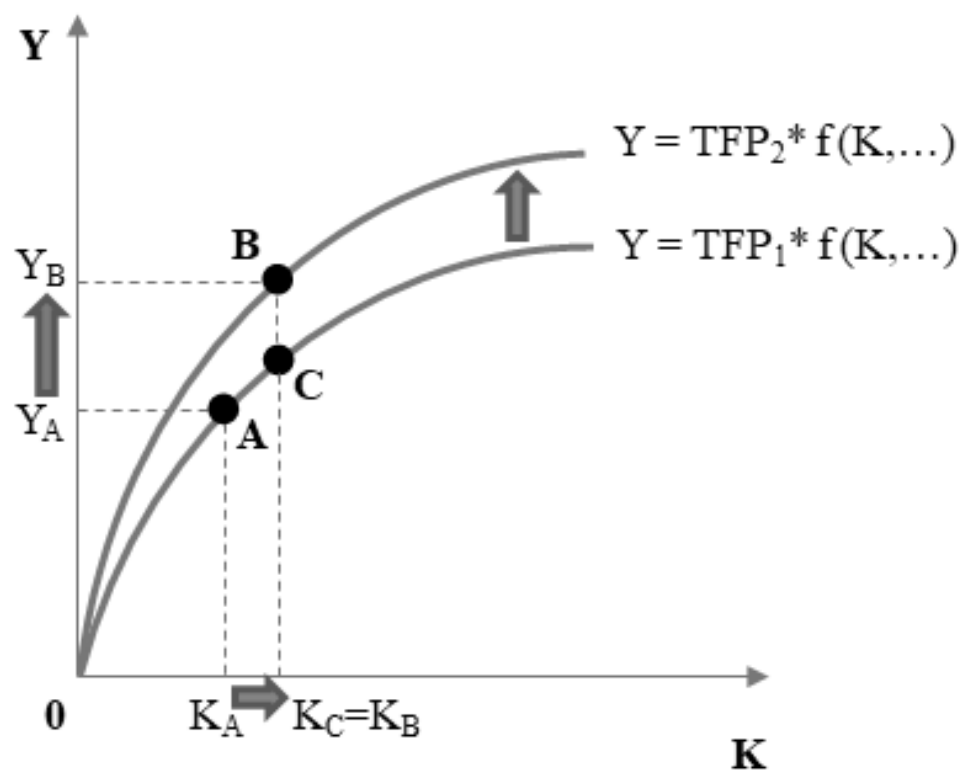

**Figure 2:** AI Emergence as Capital Growth and as AI-Driven TFP Growth in the Aggregate Production Function

**Note:** *The figure illustrates two channels through which Artificial Intelligence (AI) may increase aggregate output, including an increase in output resulting from the accumulation of AI as an additional form of capital (movement along the original curve from A to C), and an upward shift in the aggregate production function from the lower curve to the upper curve (from C to B) represents AI-driven growth in output due to growth in total factor productivity (TFP).*

The movement from point A to point C along the initial production function captures the direct contribution of AI as an additional form of capital. As firms invest in AI systems, computational infrastructure, algorithms, and related technologies, the economy's effective capital stock increases from ($K_A$) to ($K_C$). Holding the TFP level constant, this increase in capital raises output from ($Y_A$) to ($Y_C$, which will fall somewhere between $Y_A$ and $Y_B$). This effect on output growth is due to the contribution of capital accumulation to economic growth.[7]

However, AI contributes to economic growth through more than capital accumulation and capital deepening alone. AI may also improve the efficiency with which existing labor, capital, knowledge, and organizational resources are utilized. Such improvements are captured by TFP. The upward shift of the aggregate production function from the lower curve to the upper curve represents this AI-driven increase in TFP. At the same level of capital stock ($K_B = K_C$), output increases further from point C on the original production function to point B on the higher production function. Thus, the vertical distance between points C and B measures the additional output generated through AI-induced productivity improvements beyond the direct effects of capital accumulation.

Taken together, the figure illustrates that AI may influence economic growth through two complementary channels. First, AI may function as a new form of capital that expands productive capacity. Second, AI may enhance total factor productivity by improving efficiency, innovation, decision-making, and a better and more optimal utilization of existing resources. Therefore, the total contribution of AI to economic growth may exceed the effects traditionally associated with capital accumulation alone.

In its simplest form, productivity refers to the amount of output produced per unit of input (i.e. Output/Input). Economists commonly distinguish among labor productivity (Y/L), capital productivity (Y/K), and Total Factor Productivity (TFP).

Labor productivity measures output per worker (Y/L) or per hour worked (Y/N, where N denotes the number of hours worked). Capital productivity measures output generated by each unit of capital goods (Y/K). TFP refers to the existing level of overall productivity in an economy, and its growth captures the portion of increase in output that cannot be explained solely by increases in the quantities of labor or capital inputs. TFP is influenced by not only labor productivity and capital productivity, but it is also influenced by other endogenous sources of economic growth, such as human capital development, continuous knowledge accumulation, and technological innovation. As such, it is often interpreted as a measure of technological progress. Artificial Intelligence affects all three types of productivity

[7]. Readers interested in a more comprehensive treatment of the Growth Accounting model and the relationship between increased labor, capital accumulation, productivity growth, and economic growth are referred to Zeytoon-Nejad (2025).

measure described above in the ways that are to be explained below.

First of all, AI increases labor productivity (Y/L) by enabling workers to perform tasks more efficiently. Employees equipped with AI-assisted tools can often complete activities that previously required substantially more time and effort. Second, AI increases the productivity of capital (Y/K) by improving capital goods utilization, maintenance scheduling, inventory management, logistics coordination, and production planning. Third, and perhaps most importantly, AI contributes to TFP growth by a more optimal re-assignment of tasks, re-arrangement of production processes, and the ability of Large Language Models (LLMs) and other AI systems to leverage vast amounts of accumulated knowledge and information. These capabilities improve the efficiency with which labor, capital, and information are combined throughout the economy, thereby increasing output without requiring increases in the quantities of factor inputs.[8]

In sum, AI should be viewed not merely as another productive input but as a technology that enhances productivity through a number of ways simultaneously, and as such, it serves as a multi-channel engine of productivity growth that operates through automation, augmentation, optimization, prediction, and innovation.

### 3.3 The Five Productivity Channels of AI

The productivity effects of AI operate through a multitude of distinct channels:

***1. Automation:*** The most visible effect of AI on productivity is through automation. By performing tasks previously carried out by humans, AI reduces labor requirements for a given level of output. Administrative processing, customer support, data entry, document review, quality control, forecasting, and numerous other activities can increasingly be automated. Automation reduces costs and increases output per worker, thereby contributing directly to productivity growth.

***2. Augmentation:*** A second effect of AI on productivity level is through augmentation. In many cases, AI does not replace workers but instead serves as a complement and enhances workers' capabilities. Physicians assisted by diagnostic algorithms, lawyers supported by document-analysis tools, engineers using AI-assisted design software, and researchers employing AI-based analytical systems can often achieve higher levels of productivity than would otherwise be possible. In a sense, AI empowers all these types of workers, professionals, and experts, and acts as a form of cognitive augmentation that amplifies their expertise, improves decision-making, and enhances overall productivity and output.

The economic significance of augmentation may ultimately exceed that of automation because it combines machine capabilities with human judgment, creativity, and contextual understanding.[9]

***3. Optimization:*** AI can improve economic performance by optimizing resource allocation. Machine-learning systems can identify patterns within enormous datasets that would be difficult or impossible for humans to detect. These capabilities

[8]. Moreover, AI may influence the growth process itself by accelerating innovation and knowledge creation. If AI substantially increases the rate at which new technologies are discovered and implemented, it may generate a self-reinforcing cycle of technological advancement and economic growth. For this reason, one can view AI not merely as another growth-enhancing technology but as a technology capable of increasing the rate of technological progress itself.

[9]. In many occupations, the greatest economic value arises not from routine execution alone but from the ability to interpret information, exercise judgment under uncertainty, take responsibility, communicate with others, adapt to changing circumstances, and generate novel ideas. AI systems often excel at processing information, identifying patterns, and generating recommendations, while humans retain comparative advantages in judgment, intuition, ethical reasoning, interpersonal interaction, and contextual decision-making. Hence, the combination of human and artificial intelligence may prove more productive than either operating independently. If so, the largest economic gains from AI may arise not from 'substituting and replacing' workers, but from 'complementing and empowering' workers to become substantially more productive than they otherwise could be.

improve inventory management, logistics coordination, energy consumption, pricing strategies, supply-chain operations, and production scheduling. Optimization reduces waste and increases efficiency throughout the production process.

AI may also improve economic performance by enhancing individual decision-making. By providing rapid access to vast amounts of information through web searching, web scraping, and the synthesis of information from multiple sources, AI enables individuals to conduct more comprehensive and accurate cost-benefit analyses when evaluating alternative courses of action. Rather than relying on limited information or incurring substantial search costs, decision makers can access relevant data, compare competing alternatives, evaluate trade-offs, and obtain evidence-based recommendations with considerably less time and effort.

In doing so, AI reduces the cognitive burden associated with processing complex information, lowers the cost of acquiring and analyzing information, and expands the set of feasible alternatives that individuals can realistically consider. These capabilities facilitate more informed, rational, and economically efficient decision-making by improving both the quantity and quality of information available to consumers, workers, managers, entrepreneurs, and policymakers.

***4. Prediction:*** Many economic decisions are dynamic and temporal by nature, meaning that some of their possible costs or benefits are to take place in the future. Making such decisions rationally and optimally depend fundamentally upon predictions. Firms must forecast demand, investors must estimate future returns, lenders must assess credit risk, and governments must anticipate future economic conditions. AI substantially improves predictive capabilities by processing vast amounts of information and identifying relationships that may otherwise remain hidden. Improved prediction reduces uncertainty and enables more rational and optimal decision-making and more efficient allocation of resources throughout the economy when faced with dynamic economic and business decisions.

***5. Innovation:*** Perhaps the most transformative productivity-growth channel is through innovation. AI increasingly assists in scientific discovery, product design, engineering development, pharmaceutical research, and technological experimentation. Historically, innovation has been constrained by human cognitive limitations. AI expands society’s capacity to generate and evaluate new ideas, potentially accelerating the rate of technological progress itself. This effect is particularly significant because innovation influences not only current productivity but also future productivity growth, and as such, it has a compounding effect on productivity growth.

### 3.4 AI, Creative Destruction, and Long-Run Growth

The productivity effects of AI must also be understood within the broader framework of creative destruction. As discussed in the previous section, economic progress emerges not only through efficiency improvements within existing industries but also through the continual replacement of older technologies, products, occupations, and business models by newer and more productive alternatives.

AI is accelerating this process. Firms that successfully adopt AI may gain substantial competitive advantages over rivals that fail to adapt.[10] Entire industries may experience restructuring as AI-enabled firms develop superior products, lower-cost production methods, and more efficient organizational structures.

[10]. For a detailed discussion of how firms can strategically leverage Artificial Intelligence to maximize profits, improve both the supply and demand sides of their operations, and utilize AI agents and systems through four distinct pathways to enhance firm performance, see Zeytoon-Nejad (2026b), *AI and Profit: Artificial Intelligence and the Bottom Line of the Firm*.

Although such transformations may generate short-run disruptions, they also facilitate the reallocation of labor, capital, and entrepreneurial talent toward more productive activities. Historically, this process of creative destruction has been one of the primary drivers of long-run economic growth.

Thus, a significant portion of AI's economic impact may arise not from improving existing activities but from creating entirely new forms of economic activity that do not yet exist, and this is separate from incremental improvements in existing technologies.

### 3.5 Limitations and Measurement Challenges

Despite its enormous potential, measuring the productivity effects of AI presents several challenges. Many benefits generated by AI may not be fully captured in conventional economic models. For example, improvements in quality, convenience, personalization, and information access often create substantial consumer value but such improvements, say increased quality for example, are hard to measure.

Furthermore, productivity gains may initially appear modest despite substantial technological progress. Similar patterns were observed during previous technological revolutions, including electrification and the computer age. Firms often require years or even decades to redesign organizational structures and production processes sufficiently to realize the full benefits of transformative technologies. Consequently, the ultimate economic impact of AI may be substantially larger than current productivity statistics may suggest.

In short, AI enhances the efficiency with which labor, capital, information, and knowledge are utilized throughout the economy through automation, augmentation, optimization, prediction, and innovation. More importantly, AI may influence not only the level of productivity but also the rate at which productivity itself improves. By accelerating innovation and intensifying the process of creative destruction, AI has the potential to reshape the long-run growth trajectory of economies around the world.

The next section examines a more controversial dimension of this transformation: the impact of Artificial Intelligence on labor markets, employment, wages, and the future of work.

## 4. Artificial Intelligence, Labor Markets, and the Future of Work

**(*Will Artificial Intelligence replace human workers or enhance their productivity?*)**

### 4.1 Introduction

Few aspects of Artificial Intelligence have generated more economic debate than its potential effects on labor markets. Throughout history, major technological innovations have altered the nature of work, displaced certain occupations, created new ones, and transformed the skills demanded by employers. The rise of AI has reignited these debates because, unlike many previous technologies, AI increasingly affects cognitive tasks traditionally performed by educated and highly skilled workers.

The central question is not whether AI will change labor markets—it already is—but rather how and in what direction those changes will unfold and what their long-run consequences will be. Will AI primarily 'replace' workers or 'complement' them? Will 'new occupations' emerge rapidly enough to offset 'displaced jobs'? Will productivity gains translate into higher wages, or will they accrue primarily to the owners of AI-related capital? These questions lie at the heart of the economics of AI and will significantly influence the distribution of the technology's benefits throughout society.

Accordingly, this section is to examine the effects of AI on employment, wages, occupational structure, and income distribution.

### 4.2 The Historical Perspective: Technology and Employment

Concerns regarding technological unemployment are not new. The Industrial Revolution displaced many traditional artisans and craftsmen. Mechanization reduced demand for agricultural labor. Automation transformed manufacturing employment. Computers altered clerical and administrative occupations.

Yet despite these disruptions, long-run employment generally continued to expand because technological progress simultaneously increased productivity, reduced costs, expanded output, created new industries, and generated demand for new forms of labor. In other words, although many existing jobs at the time disappeared as a result of technological change, the dynamics outlined above (i.e., productivity gains, cost reductions, market expansions, and emergence of entirely new industries) ultimately created even more employment opportunities, causing total employment and living standards to rise rather than fall over the long run.[11]

This historical experience suggests an important distinction. Technology often destroys specific jobs, but it does not necessarily reduce the total demand for labor permanently. Instead, it changes the composition of labor demand. Whether AI follows a similar pattern remains one of the most important questions in modern economics.

### 4.3 Automation: The Labor-Substitution Effect

The most widely discussed effect of AI is automation. AI systems can increasingly perform tasks that were previously carried out by human workers, including customer service, data entry, administrative processing, bookkeeping, translation, document review, quality control, forecasting, basic coding, and content generation. When AI performs these tasks at lower cost or higher quality and accuracy, firms tend to substitute AI for labor as they pursue their natural objective of profit maximization.

From an economic perspective, this represents the labor-substitution effect. The magnitude of this effect depends upon several factors, including the cost of AI relative to labor (i.e., the relative cost), the quality of AI-generated output (i.e., reliance), regulatory constraints, consumer preferences, and the degree of task routinization.

Occupations characterized by repetitive and predictable tasks are generally more vulnerable to automation than occupations requiring complex interpersonal interactions, creativity, judgment, leadership, or emotional intelligence.

### 4.4 Augmentation: The Labor-Complementarity Effect

Although automation receives considerable attention, it represents only part of the story. In many cases, AI functions as a complement to labor rather than as a substitute for it. Professionals increasingly use AI systems to improve decision-making, increase efficiency, and enhance productivity. Physicians employ diagnostic algorithms. Engineers utilize AI-assisted design tools. Financial analysts rely on predictive models. Researchers leverage AI to analyze large datasets and generate insights. In these contexts, AI functions as a productivity-enhancing *complement* to human labor rather than a *substitute* for it.

Instead of replacing workers, AI often acts as a form of cognitive augmentation that amplifies human expertise, improves decision-making, and enhances overall productivity. Therefore, many occupations may experience transformation rather than elimination.

[11]. Appendix A provides a more detailed analytical treatment of these labor-market dynamics by developing a mathematical framework and accompanying visual representations of the static and dynamic employment effects of Artificial Intelligence. Specifically, it illustrates how AI affects the total number of jobs in the economy by distinguishing between employment losses within existing firms resulting from labor substitution and employment gains arising from the creation of new firms, new industries, new occupations, and new entrepreneurial opportunities. By integrating these components into a unified framework, the appendix demonstrates that the ultimate employment effect of AI depends on the relative magnitudes of job destruction and job creation, thereby providing a complete visualization of the process of AI-driven creative destruction in the labor market.

### 4.5 The Reallocation Effect

A critical lesson from economic history is that technological change rarely affects all sectors equally. As AI automates certain activities, labor and capital are reallocated toward new activities that become economically viable and available because of technological progress. This process reflects the broader dynamics of creative destruction which was discussed in the previous sections. Some occupations decline, while others expand.

Historically, societies have continually generated new occupations that were previously unimaginable. Software developers, data scientists, social media managers, game designers, stand-up comedians active in social media, cloud-computing specialists, and cybersecurity professionals are examples of occupations that barely existed a few decades ago.

Similarly, AI is likely to generate entirely new categories of employment that cannot yet be fully anticipated. The challenge lies not in the eventual creation of new opportunities but in the transitional adjustment process and time required to move workers from declining sectors into expanding ones.

### 4.6 Skill-Biased Technological Change

One of the most important labor-market implications of AI is its potential to intensify skill-biased technological change. Historically, many technological innovations have increased the relative demand for highly educated and highly skilled workers. AI may strengthen this tendency. Workers capable of effectively utilizing AI systems may experience substantial demand for their skills, while workers whose tasks can be readily automated may face declining demand for their skills.

As a result, wage differentials between highly skilled and less-skilled workers may widen. The labor market may increasingly reward analytical skills, creativity, problem-solving abilities, adaptability, technical literacy, and interpersonal capabilities. On the contrary, occupations heavily dependent upon routine and predictable activities may experience increasing competitive pressures.

### 4.7 AI and Income Distribution

The effects of AI extend beyond employment and wages to the broader distribution of income. Because AI can function simultaneously as capital and labor, it may alter the historical balance between labor income and capital income.

Traditionally, workers supplied labor while investors supplied capital. AI blurs this distinction by enabling capital assets and equipment to perform tasks previously undertaken by labor. As a result, a larger share of economic output may accrue to owners of AI systems, intellectual property, computational infrastructure, and complementary digital assets.

This creates the possibility of increasing income inequality even in the presence of substantial productivity growth. The distributional consequences of AI therefore depend not only on how much wealth is created but also on who creates and owns the technologies responsible for creating the wealth, who adopts and leverages AI effectively, and whose skills and occupations are displaced by AI specialists or AI systems.

### 4.8 The Future of Work

Contrary to many popular narratives, the future of work is unlikely to be characterized by the complete elimination of human labor. Rather, work itself is likely to evolve. Routine activities will increasingly be automated. Human labor will become concentrated in tasks involving judgment, creativity, leadership, entrepreneurship, interpersonal interaction, ethical reasoning, and complex problem-solving. Many occupations will become hybrid occupations in which humans and AI systems collaborate to achieve outcomes neither could accomplish as effectively alone.

The future of work may therefore be defined less by human-versus-machine competition and more by human-machine cooperation. Workers who successfully adapt to this environment may experience substantial productivity gains, while those who fail to adapt may encounter increasing labor-market challenges.

In conclusion, Artificial Intelligence is transforming labor markets through multiple channels, including labor substitution, labor augmentation, the creation of new occupations, the reallocation of labor across tasks and industries, skill-biased technological change, and the increasing importance of AI-related human capital. While AI will undoubtedly displace certain tasks and occupations, it is also likely to create new opportunities, new industries, and new forms of work.

The ultimate impact of AI on employment and wages will depend not only on technological capabilities but also on the adaptability of workers, firms, educational institutions, and public policy. The historical record suggests that technological progress has generally expanded economic opportunities over the long run, but the transition process can be disruptive, lengthy, and uneven. Consequently, the central labor-market challenge posed by AI is not simply 'technological adoption' but 'economic adaptation' and receptivity to change. Economies that successfully facilitate this adaptation within a reasonable timeframe are likely to capture a larger share of AI's productivity gains while minimizing its social costs.

## 5. Artificial Intelligence, Market Structure, Competition, and Industrial Organization

**(*Will Artificial Intelligence promote 'competition' or increase 'market concentration'?*)**

### 5.1 Introduction

The economic consequences of Artificial Intelligence extend far beyond productivity growth and labor-market transformation. AI is also reshaping the structure of markets and industries, the nature of competition, and the organization of firms. As AI technologies become increasingly integrated into production processes, decision-making systems, and consumer-facing applications, they are altering the competitive landscape of modern economies.

Historically, technological innovation has often produced conflicting effects on market structure. On the one hand, innovation can lower barriers to entry, create new entrepreneurial opportunities, and intensify competition. On the other hand, successful innovations can generate economies of scale, network effects, and first-mover advantages that strengthen the market power of dominant firms, and thereby undermine competition. Artificial Intelligence exhibits both tendencies simultaneously.

This section is to examine how AI can influence market concentration, barriers to entry, economies of scale, network effects, entrepreneurship, and industrial organization.

### 5.2 AI and Economies of Scale

One of the defining economic characteristics of AI is its extraordinary scalability. Traditional production processes often require proportional increases in labor and capital as output expands. AI systems differ because, once developed and deployed, they can frequently serve large numbers of users at relatively low marginal cost.[12]

Training an advanced AI model may require substantial upfront investments in research,

[12]. Recent developments in AI-enabled business models provide striking examples of the scalability discussed here. For example, a 2026 New York Times report described Medvi, an AI-enabled telehealth startup that was reportedly on track to generate approximately $1.8 billion in annual revenue while employing only two individuals (Griffith, 2026). Such cases illustrate how AI can substantially decouple output and revenue generation from traditional labor requirements, allowing firms to serve large numbers of customers with minimal increases in employment and operating costs. While such examples remain exceptional rather than representative of the broader economy, they demonstrate the unprecedented scalability that AI technologies can introduce into modern production processes and business organizations.

computational infrastructure, and data acquisition. However, once developed, the same model can often be utilized repeatedly across numerous applications and customers.[13]

This cost structure creates significant economies of scale. As output expands, average costs decline due to economies of scale, providing larger firms with important competitive advantages. Consequently, AI-intensive industries may naturally tend toward greater concentration than industries characterized by conventional production technologies.

Figure 3 exhibits this process and characteristics in a visual manner.

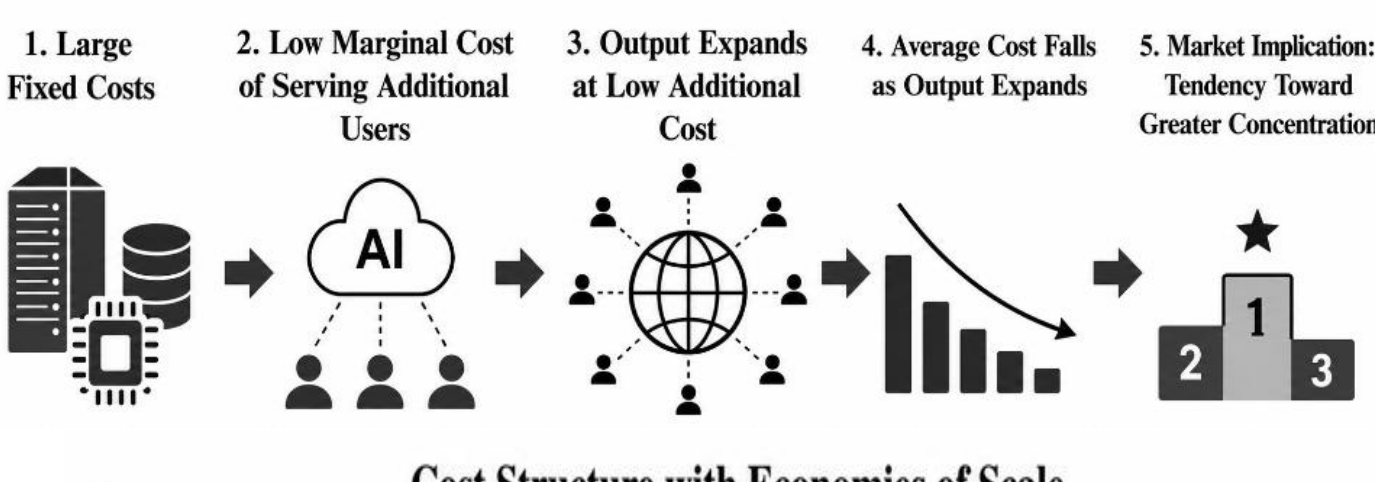

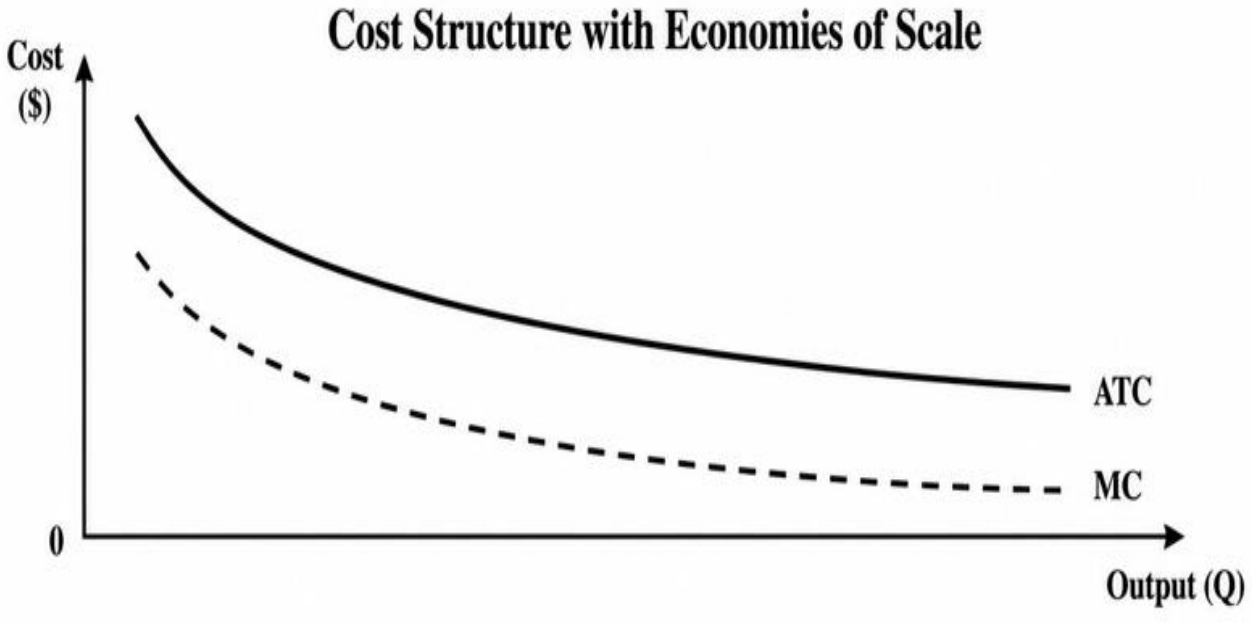

**Figure 3:** Economies of Scale in AI Production and Their Implications for Cost Structure and Market Concentration

[13]. From the perspective of the typology of goods in economics, provision of AI systems exhibits several characteristics of a club good. Once an AI model has been developed, it can often serve a large number of users simultaneously at very low marginal cost, making it largely non-rival until computational capacity becomes congested. At the same time, access is excludable through subscriptions, licensing agreements, or application programming interfaces (APIs). Unlike many traditional club goods that are often associated with natural monopolies or exclusive membership organizations, however, AI services are generally supplied by multiple competing firms operating in increasingly oligopolistic—and in some segments, highly competitive—markets. Thus, AI may be viewed as possessing club-good characteristics on the demand side while being supplied through oligopolistic or competitive market structures on the supply side.

### 5.3 Data as a Strategic Asset

In the AI economy, data has become an increasingly important productive resource. Machine-learning systems rely heavily on data for training, refinement, and performance improvement. Firms possessing larger and higher-quality datasets often enjoy significant advantages in developing and improving AI systems.

This creates a self-reinforcing dynamic. More users generate more data. More data improves AI performance. Better performance of the firm attracts additional users. Additional users generate even more data. Figure 4 exhibits this self-reinforcing feedback loop.

This feedback mechanism can create substantial barriers to entry for potential competitors. Therefore, unlike traditional physical assets, valuable datasets often cannot be easily replicated or purchased in competitive markets. Hence, data ownership may become an increasingly important source of competitive advantage in AI-intensive industries.

### 5.4 Network Effects and Platform Dominance

Artificial Intelligence also strengthens network effects. A network effect exists when the value of a product or service increases as the number of users increases. Social media platforms, digital marketplaces, payment systems, and communication networks all exhibit network effects. AI can amplify these effects.

As platforms accumulate larger user bases, they gain access to larger datasets, which improve the performance of AI algorithms. Improved performance attracts additional users, creating a virtuous cycle of growth, which is depicted in Figure 4.

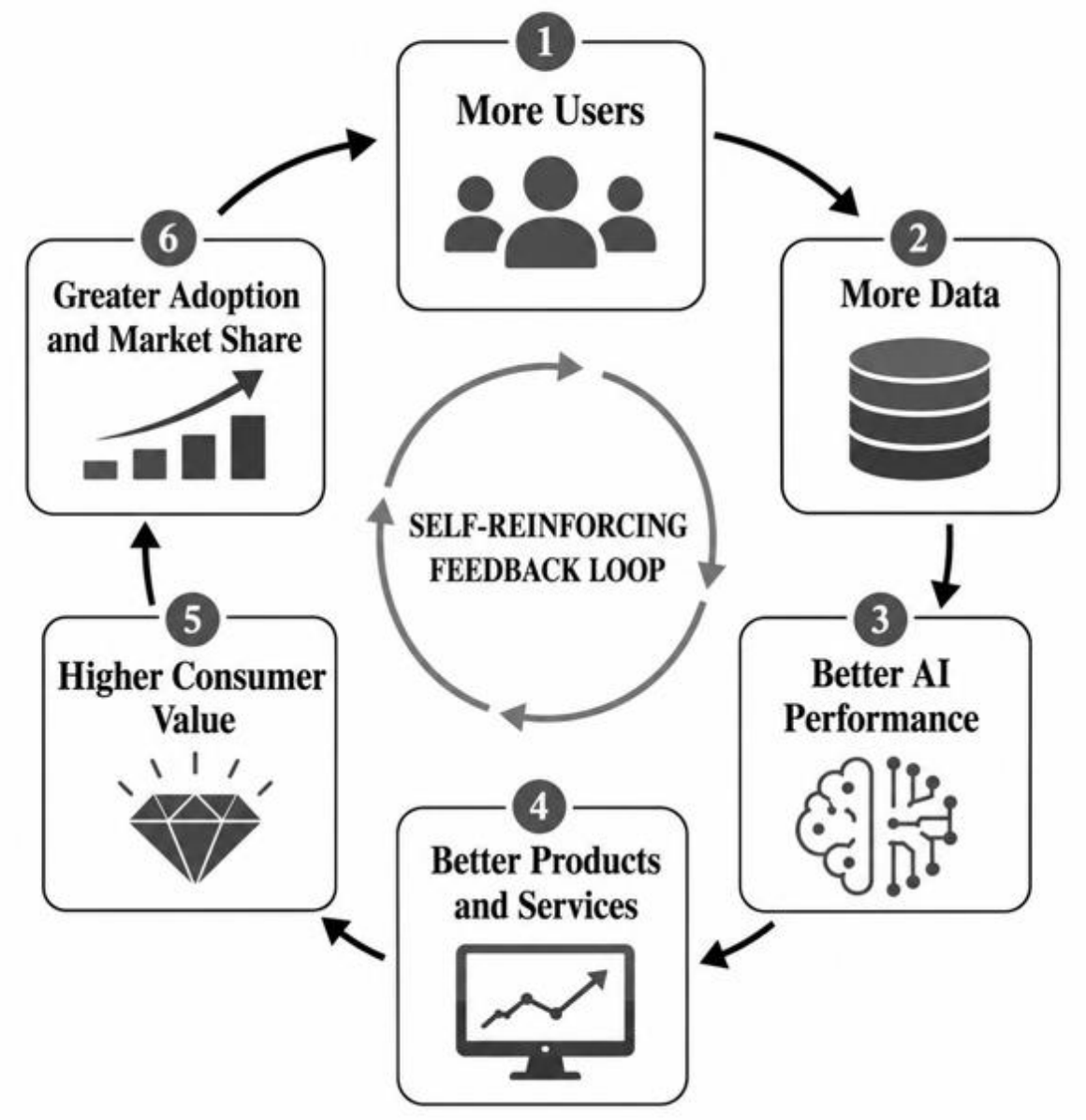

**Figure 4:** The Data–AI Feedback Loop and Data Network Effects

**Note:** *The figure illustrates the self-reinforcing relationship between user adoption, data accumulation, AI performance, and market expansion. As additional users generate more data, AI systems improve, creating greater consumer value and attracting additional users, thereby strengthening the competitive position of the firm.*

This process can generate winner-take-most market outcomes in which a relatively small number of firms capture a disproportionate share of market activity. As a result, AI may contribute to increasing concentration within certain industries, particularly those characterized by strong digital-network effects.

### 5.5 AI and Barriers to Entry

The effect of AI on barriers to entry is more complex than it initially appears. On the one hand, advanced AI development often requires large datasets, significant computational resources, specialized expertise, and substantial financial investment. These requirements may raise barriers to entry and strengthen incumbent firms.

On the other hand, many AI tools are becoming increasingly accessible and affordable. Small businesses and entrepreneurs can now access powerful AI capabilities through cloud-based platforms and subscription services without developing their own systems from scratch.

Thus, AI simultaneously increases and decreases barriers to entry depending upon the specific industry and application under consideration. The net effect depends on the relative strength of these opposing mechanisms. '*The concentration effect*' arises from the substantial fixed costs of developing advanced AI systems and the competitive advantages associated with large datasets and computational resources. '*The democratization effect*' arises from the increasing availability of low-cost AI tools that enable smaller firms to access capabilities that were previously available only to large organizations.

Whether AI ultimately increases or decreases market concentration therefore depends on which of these two effects dominates. If the former dominates, industries may become more concentrated and entry barriers may rise. If the latter dominates, competition may intensify as new firms and entrepreneurs gain access to powerful productive capabilities at relatively low cost.

### 5.6 AI and Entrepreneurship

Although discussions of AI frequently focus on large technology firms, AI also creates significant entrepreneurial opportunities. Historically, technological revolutions have often generated entirely new industries and business models. Artificial Intelligence is likely to follow a similar pattern. Entrepreneurs increasingly utilize AI to develop new products and services, automate business operations, reduce startup costs, enhance customer engagement, improve market analysis, and accelerate innovation.

By lowering the cost of information processing and decision-making, AI may enable smaller firms to compete more effectively with larger organizations. Therefore, while AI may increase concentration in some sectors, it may simultaneously promote entrepreneurial dynamism in others.

### 5.7 AI and the Boundaries of the Firm

Artificial Intelligence may also influence the size and structure of firms themselves. According to transaction-cost economics, firms exist because internal coordination is sometimes less costly than market transactions.

AI reduces information costs, monitoring costs, communication costs, and coordination costs. As these costs decline, firms may reorganize their operations, automate managerial functions, and redesign organizational structures. Some firms may become larger due to scale advantages, while others may become smaller and more flexible because AI reduces the costs of coordinating complex activities. Therefore, AI has the potential to reshape not only industries but also the internal organization of firms.

### 5.8 Creative Destruction and Competitive Dynamics

The relationship between AI and competition must also be viewed through the lens of creative destruction. As discussed earlier, AI represents one of the latest waves of innovation within the broader evolutionary process of capitalism. New technologies often disrupt existing firms and industries. Incumbent firms that fail to adapt may lose market share or disappear entirely, while innovative entrants may emerge as new market leaders.

The history of economic growth and market development demonstrates that market dominance is rarely permanent. Many firms that once appeared unassailable were eventually displaced by new technologies and business models. As a result, AI simultaneously strengthens incumbent advantages and creates opportunities for disruptive entrants. The ultimate effect on competition will depend largely upon the balance between these two opposing forces.

### 5.9 Market Power and Regulatory Challenges

The possibility of increasing power concentration raises important public-policy questions. If AI contributes to an excessive market power that is not legitimate (say, not due to providing a uniquely valuable, differentiated product) and is therefore concerning, policymakers may face concerns regarding higher barriers to entry, reduced competition, and other consequential changes occurring to consumer welfare, data ownership, privacy, and innovation incentives.

However, regulatory intervention must be approached carefully. Excessive restrictions on AI development may reduce innovation and slow down productivity growth, which in turn can negatively influence living standards. Insufficient oversight, on the other hand, may permit the emergence of anti-competitive practices and the creation of excessive market concentration. The real challenge for policymakers is therefore not to suppress AI-driven innovation but to ensure that markets remain contestable, competitive, and open to entrepreneurial entry.

In conclusion, Artificial Intelligence is reshaping industrial organization through economies of scale, data advantages, network effects, changing barriers to entry, and new forms of entrepreneurship. These forces create both opportunities and risks. AI may simultaneously increase market concentration in some sectors while stimulating innovation and entrepreneurial dynamism in others. The ultimate competitive landscape of the AI economy will depend upon how these opposing forces evolve over time. While AI creates powerful incentives toward scale and concentration, it also fuels the process of creative destruction through which new firms and technologies continually challenge established market leaders.

The next section broadens the analysis beyond firms and markets to examine the welfare implications of AI, including its effects on consumers, equality, social welfare, and economic policy.

## 6. Artificial Intelligence, Welfare, Prosperity, and Public Policy

***(Does AI ultimately make society better off?)***

### 6.1 Introduction

The ultimate objective of economic activity is not production for its own sake but the improvement of human well-being. Productivity, innovation, capital accumulation, and economic growth are important primarily because they enable societies to generate greater prosperity, expand opportunities, and improve living standards. Hence, the most important question surrounding Artificial Intelligence is not merely whether it increases output, but whether it improves overall social welfare.

The preceding sections have demonstrated that AI possesses the potential to accelerate innovation, enhance productivity, transform labor markets, and reshape industrial organization. However, these economic effects do not automatically translate into improvements in social welfare for all. The welfare implications of AI depend upon both the magnitude of the benefits generated and the manner in which those benefits are distributed throughout society.

This section examines the welfare effects of AI from the perspectives of consumers, producers, workers, governments, and society as a whole.

### 6.2 Consumer Welfare and the Expansion of Economic Value

One of the most direct benefits of AI arises through improvements in consumer welfare. By reducing production costs, improving product quality, enhancing personalization and customization, and increasing the availability of information, AI enables consumers to obtain greater value from economic activity, which in turn can potentially translate into larger consumer surplus for demanders, too.

Examples of ways through which consumers can benefit from AI include more accurate search results, personalized recommendations, improved healthcare diagnostics, faster customer service, better navigation systems, enhanced educational tools, and more efficient financial services. Many of these benefits generate substantial value even when they are not fully reflected in customary economic metrics. As a result, the welfare gains generated by AI may exceed what is observable through conventional measures.

### 6.3 Producer Welfare and Cost Efficiency

Firms also benefit from AI through increased efficiency and profitability, which in turn can potentially translate into greater producer surplus for suppliers, too. In fact, AI reduces information-processing costs, improves resource allocation, enhances forecasting accuracy, optimizes production processes, and facilitates innovation for firms. These improvements increase productive efficiency and expand (i.e., shifts outward) the economy’s production possibilities frontier.

From a welfare perspective, increased efficiency allows society to generate more output from a given quantity of resources, thereby increasing the total amount of wealth available for consumption, investment, and future growth. Therefore, AI may contribute significantly to long-run prosperity through its effects on productive efficiency.

### 6.4 The Distributional Effects

Although AI tends to increase aggregate welfare, the benefits generated by AI are unlikely to be realized simultaneously for all. Additionally, the benefits are unlikely to be to the same extent for all individuals, firms, and industries. The owners of AI-related capital, holders of intellectual property, highly skilled workers, and large technology firms tend to gain more at early stages. Additionally, the firms and businesses that acquire AI-related skills, adopt AI technologies, invest in AI-enabled capital, or successfully integrate AI into their production processes earlier are likely to capture a larger share of the initial gains. Early adopters often benefit from early-learning advantages, productivity improvements, market expansion opportunities, and first-mover advantages that may not be immediately available to later adopters.[14]

At the same time, some workers, firms, and industries may face adjustment costs, temporary layoffs, declining demand for their services, displacements and frictional unemployment as economic activity is reorganized around new technologies and new production methods. An increase in aggregate welfare does not automatically imply that all individuals benefit equally or at the same time. Rather, the gains from technological progress tend to accrue first to those who adapt most rapidly to changing economic conditions.

In this sense, one of the most important lessons of the AI economy is that those who move first and adapt sooner gain more. As AI diffuses throughout the economy, competitive advantages are likely to emerge for individuals and organizations that acquire AI-related knowledge, develop complementary skills, and learn how to leverage AI more quickly and more effectively. Over time, however, as adoption becomes more widespread, many of these gains will diffuse more broadly throughout society. In this sense, a process analogous to the so-called 'trickle-down' effect may occur, as the widespread diffusion of AI, competitive market forces, and declining costs allow an increasingly broader segment of society to benefit from the technology, and then a broader range of population will benefit from AI on net, much as occurred with earlier waves of technological innovation.

[14]. Appendix B develops this idea further by presenting a conceptual framework that categorizes firms according to their likely relationship with Artificial Intelligence and the process of creative destruction. Specifically, it distinguishes between AI beneficiaries, AI adopters and business-model transformers, AI-augmented firms, AI-disrupted firms, and potential AI casualties, illustrating the diverse ways in which AI may reshape industries and competitive dynamics.

### 6.5 Dynamic Welfare and Creative Destruction: Short-Run Pains and Long-Run Gains

The welfare implications of AI should also be analyzed dynamically rather than statically. Many technological innovations generate short-run adjustment costs while simultaneously producing long-run benefits. Workers may need to acquire new skills. Firms may need to alter business models. Industries may undergo substantial restructuring. These transitional costs are often highly visible and politically salient. However, they must be weighed against the long-run benefits generated by increased innovation, productivity, and economic growth.

The history of technological progress suggests that societies have repeatedly accepted temporary disruptions because the resulting improvements in living standards ultimately proved substantially larger than the associated costs. AI is likely to represent another example of this broader process of creative destruction.

### 6.6 AI and the Expansion of Human Capabilities

Beyond its effects on output and income, AI may improve welfare by expanding human capabilities and empowering people. Individuals increasingly utilize AI to access information, acquire new skills, improve decision-making, enhance creativity, increase productivity, and solve complex problems. In this sense, AI functions not only as a productive technology but also as a capability-enhancing, empowering technology.

As a result, its welfare contribution in the long run can extend beyond conventional economic metrics and include improvements in knowledge, opportunity, convenience, and personal empowerment as well, which are achievements that are harder to measure.

### 6.7 AI, Government Revenue, and the Provision of Public Goods

Artificial Intelligence may also generate important welfare benefits for governments and the public sector. To the extent that AI increases productivity, reduces production costs, stimulates innovation, and expands economic activity, it can enlarge the tax base from which governments derive revenue.

From a microeconomic perspective, AI often increases the marginal value to consumers of products generated by productive activities while simultaneously reducing the marginal cost to producers of producing goods and services. These effects tend to shift supply and demand conditions favorably, lowering marginal costs and increasing marginal value, and thereby increase the quantity of goods and services exchanged in markets. As output and exchange expand and transactions increase, total economic surplus—the difference between marginal value and marginal cost, which is equal to the sum of consumer surplus and producer surplus—also tends to increase.

The expansion of total surplus has important implications for public finance. Holding tax rates constant, a larger volume of production, consumption, income, profits, and market transactions generally results in higher tax revenues. Therefore, AI-driven economic growth may enable governments to collect greater tax revenues without necessarily increasing tax rates. In this sense, AI may contribute not only to private prosperity but also to the fiscal capacity of the state.

Greater fiscal capacity may, in turn, allow governments to expand the provision of public goods and public services, including transportation infrastructure, education, scientific research, public health systems, law enforcement, judicial institutions, environmental protection, and national defense. Such investments can further enhance economic performance by improving the institutional and physical foundations upon which private economic activity depends, as public goods and private goods can be complements to some degree.

In fact, many public goods function as complementary inputs into private production. Infrastructure facilitates commerce and logistics. Educational systems improve human capital. Scientific research generates knowledge spillovers that support innovation. Legal and judicial institutions reduce transaction costs and strengthen property rights. Thus, increased government capacity to provide public goods may further stimulate private-sector productivity and economic growth in turn.

At the same time, public goods often generate utility directly for citizens. Individuals benefit from safer communities, cleaner environments, better transportation networks, improved public health systems, and greater educational opportunities. Therefore, the welfare gains associated with AI may extend beyond increases in private consumption and private production to include improvements in the quantity and quality of public goods available to society.

This process creates the possibility of a virtuous cycle, which is depicted in Figure 5.

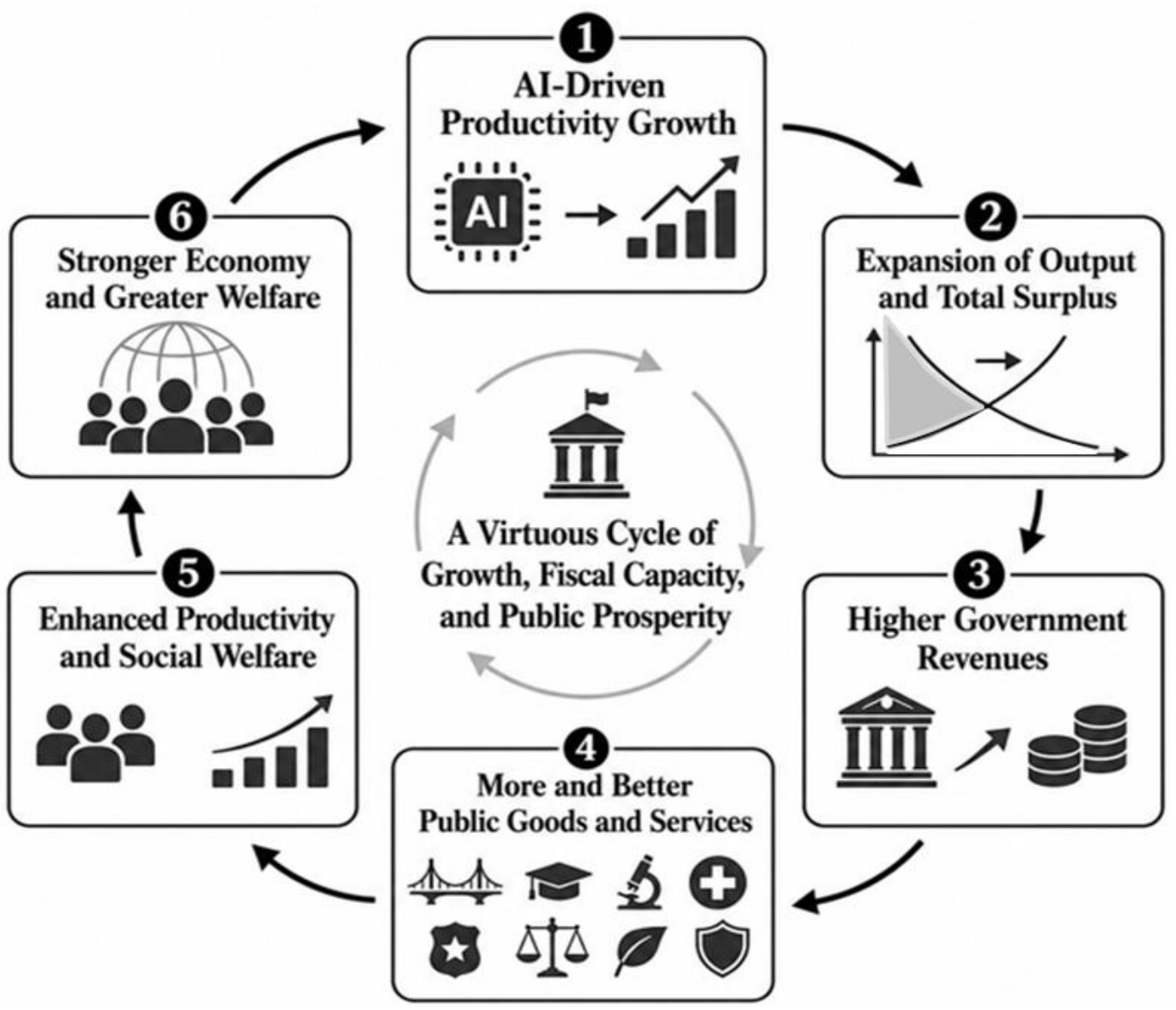

**Figure 5:** The Virtuous Cycle of AI-Driven Growth, Fiscal Capacity, and Public Prosperity

As shown in Figure 5, AI-driven productivity growth expands economic output and total surplus. Higher levels of economic activity increase government revenues. Greater revenues enable improved provision of public goods. Improved public goods further enhance private-sector productivity and social welfare. To the extent that this cycle operates successfully, the benefits of AI may be transmitted not only through markets but also through the public institutions that support economic and social development.

### 6.8 Public Policy Considerations

The welfare effects of AI depend partly on the institutional environment within which it is developed and deployed. Public policy can influence the timing, distribution, and magnitude of AI's benefits through education, job training programs (for developing an agile workforce), competition policy, intellectual-property frameworks, data governance, research support, and regulatory oversight.

The objective of policy should not be to impede technological progress. Historically, attempts to suppress innovation have often reduced long-run prosperity. Instead, policymakers should seek to maximize the benefits of AI while facilitating adaptation to the changes it creates. This requires institutions that encourage innovation, support entrepreneurship, promote competition, and help workers acquire the skills necessary to thrive in an increasingly AI-intensive economy that is being shaped and continuously evolved through the process of creative destruction.

### 6.9 The Central Welfare Question

The debate surrounding AI is often framed as a choice between optimism and pessimism. Some observers emphasize automation, displacement, and inequality. Others emphasize productivity, innovation, and prosperity. From an economic perspective, both views capture important elements of reality. AI is likely to generate substantial benefits and substantial disruptions simultaneously. The relevant question is therefore not whether AI creates winners and losers. Nearly every major technological transformation has done so. Rather, the relevant question is whether the gains generated by AI exceed the associated costs and whether institutions are capable of facilitating the necessary economic adjustments in a reasonable manner. Historically, the answer for most major technological revolutions has been affirmative.

In short, Artificial Intelligence possesses the potential to generate significant welfare gains through increased productivity, enhanced consumer value, reduced production cost, accelerated innovation, and expanded human capabilities. At the same time, it creates important challenges related to adjustment costs, market concentration, and the dynamics of distributional outcomes. The overall welfare impact of AI will ultimately depend on the balance between these opposing forces. If societies successfully harness the productive potential of AI while facilitating adaptation to its disruptive effects in a reasonable, timely manner, AI can become one of the greatest contributors to human prosperity in modern economic history. If adaptation mechanisms prove inadequate and lengthy, however, the transition may generate substantial economic and social tensions and pushbacks despite continued technological progress.

The next section aims to synthesize the major arguments and discuss the broader implications of AI for the future evolution of economic systems, markets, and human prosperity.

## 7. Artificial Intelligence and the Future of Economic Systems

**(*What kind of economic future is Artificial Intelligence likely to create?*)**

### 7.1 AI and the Evolution of Capitalism

The emergence of AI may ultimately represent a transformation not merely of production processes or labor markets, but of the economic system itself as a whole. The purpose of this section is to examine the broader implications of AI for capitalism, economic institutions, and the future organization of economic activity.

Market economies have historically evolved through successive waves of innovation and creative destruction. New technologies alter production methods, create new industries, eliminate obsolete activities, and continuously reshape economic relationships. Artificial Intelligence appears to represent one of the most significant waves of creative destruction in modern economic history.

Unlike many previous technologies, AI affects not only physical production but also information processing, decision-making, prediction, creativity, and knowledge generation. As a consequence, AI has the potential to transform a much broader range of economic activities than many earlier innovations. In this sense, AI should be viewed not as a departure from capitalism but as a continuation of capitalism's evolutionary dynamics. The market system has long served as a mechanism for experimentation, discovery, adaptation, and innovation. AI accelerates these processes by expanding the productive capabilities available to individuals, firms, and institutions.

### 7.2 The Economics of Abundance

One of the most important long-run implications of AI is the possibility of increasing economic abundance. Many economic problems originate from scarcity. Human wants are virtually unlimited, while resources are limited. Economic systems exist largely because societies must allocate scarce resources among competing, alternative uses.

By dramatically increasing productivity and reducing production costs, AI may alleviate certain forms of scarcity. Information, knowledge, design, analysis, prediction, and numerous cognitive services can increasingly be produced at very low marginal costs. As AI capabilities improve, an expanding range of economic activities may become more abundant and more accessible. This does not imply the elimination of scarcity. Physical resources, land, energy, and human time will remain limited. However, AI may substantially reduce scarcity in many knowledge-intensive activities that occupy an increasingly large share of modern economies.

### 7.3 Human Capital in an AI Economy

As AI assumes responsibility for a growing number of 'routine cognitive tasks', the relative importance of 'uniquely human capabilities' will increase. Creativity, judgment, leadership, entrepreneurship, interpersonal interaction, ethical reasoning, the ability to operate under uncertainty, and many other such human tasks are likely to remain valuable even as AI capabilities expand. Therefore, the future economy may place greater emphasis on human adaptability and continuous learning.

Educational systems may increasingly focus not merely on transmitting information but on developing analytical thinking, creativity, problem-solving skills, and the capacity to work effectively alongside AI technologies. The future of work may therefore involve less competition between humans and machines and more collaboration between them.

### 7.4 AI, Institutions, and Governance

The long-run effects of AI will depend not only on technological capabilities but also on the institutions that govern its development and deployment. Property-rights systems, legal frameworks, educational institutions, regulatory structures, and competitive markets will all influence how AI affects economic outcomes. Societies that successfully align these institutions with technological progress are likely to capture a larger share of AI's benefits. Conversely, weak institutions may hinder innovation, distort incentives, impede adaptation, or delay benefits.

As AI becomes increasingly integrated into economic decision-making, questions of transparency, fairness, accountability, competition, privacy, and governance will become progressively more important. The future economic impact of AI will therefore be shaped as much by institutional quality as by technological capability.

### 7.5 Reasons for Optimism, Reasons for Caution, and the Central Economic Question

The economics of AI contains both reasons for optimism and reasons for caution. Reasons for optimism include higher productivity, faster innovation, improved consumer welfare, greater economic efficiency, expanded opportunities for entrepreneurship, and increased prosperity and living standards. Reasons for caution include labor-market disruptions, adjustment costs, income inequality, market concentration, governance challenges, and transitional social tensions. Both perspectives contain important insights.

The historical record suggests that major technological revolutions often generate substantial disruptions during periods of transition while simultaneously producing significant long-run gains. AI is likely to follow a similar pattern. The central economic question surrounding AI is therefore whether the benefits generated by AI will outweigh the associated costs and whether economic systems can successfully adapt to the changes it creates.

From a historical perspective, market economies have repeatedly demonstrated a remarkable capacity for adaptation. New technologies have often generated fears of displacement and disruption, yet they have also produced unprecedented increases in productivity, wealth, and human welfare, and AI seems to be of the same nature and following this same pattern.

In sum, as the latest major wave of creative destruction, AI is accelerating the evolutionary dynamics that have long characterized market economies. Its ultimate impact will depend on the interaction between technological progress, market incentives, institutional quality, and societal adaptability. If these forces evolve constructively, AI may become one of the most powerful contributors to productivity growth, prosperity, and human progress in modern history. The future of AI is therefore not merely a technological question. It is fundamentally an economic question concerning how economies organize resources, incentives, institutions, and human creativity in an age of increasingly AI machines.

## 8. Policy Implications and Recommendations

**(*How can societies maximize the benefits of AI while minimizing its costs?*)**

### 8.1 Introduction

Artificial Intelligence presents policymakers with a complex challenge. On the one hand, AI possesses enormous potential to stimulate innovation, increase productivity, improve living standards, and expand economic opportunities. On the other hand, AI may generate labor-market disruptions, increase market concentration, alter income distribution, and create new governance challenges.

The objective of public policy should therefore not be to impede technological progress, nor to promote technological adoption without regard for its undesired consequences. Rather, the objective should be to create institutional conditions that maximize the benefits of AI while facilitating adaptation to the changes it generates.

This section outlines several policy principles that may help societies achieve this dual objective.

### 8.2 Preserve Innovation Incentives

The first principle is that policies should preserve incentives for innovation. Historically, technological progress has been one of the most important drivers of productivity enhancement, economic growth, and improvements in living standards. Excessively restrictive policies may discourage entrepreneurship, reduce investment, and slow the pace of innovation.

While appropriate safeguards may be necessary in certain contexts such as preserving privacy, policymakers should recognize that the long-run prosperity generated by AI depends fundamentally on continued experimentation, investment, and technological advancement. Policies that excessively burden innovation may reduce the very welfare gains that AI can make possible for societies.

### 8.3 Invest in Human Capital

Because AI is transforming labor markets, investment in human capital becomes increasingly important. Educational institutions should place greater emphasis on analytical reasoning, critical thinking, creativity, problem solving, technological literacy, adaptability, and lifelong learning.

The objective is not merely to prepare workers for existing occupations but to equip them with skills that remain valuable in a rapidly changing economic environment. In the final analysis, workforce adaptation is likely to be one of the most important determinants of whether AI generates broadly shared prosperity.

### 8.4 Promote Labor-Market Flexibility

Technological change inevitably requires economic adjustment. Labor markets that facilitate mobility between occupations, industries, and geographic regions are generally better positioned to absorb technological disruptions. Policies that reduce unnecessary barriers to labor mobility, encourage re-training, and support occupational transitions may help workers adapt more successfully to AI-driven changes.

It is important to note that the goal should not be to preserve every existing job indefinitely, but to help individuals transition toward emerging opportunities in a reasonable manner and at a reasonable pace. Public-policy instruments that can help with this issue include unemployment insurance and job training programs.

### 8.5 Maintain Competitive Markets

The benefits of AI are likely to be larger when markets remain competitive. Competition encourages innovation, limits market power, promotes the diffusion of new technologies throughout the economy, and enables the gains from AI to spread progressively throughout society as adoption becomes more widespread. Because AI may generate economies of scale, data advantages, and network effects, policymakers should remain attentive to the possibility of excessive market concentration.

At the same time, competition policy should distinguish between legitimate market dominance achieved through superior innovation and strong, valid value proposition and market dominance sustained through rent-seeking, anti-competitive behavior. In this regard, the objective should be to develop contestable markets rather than artificial limitations on successful firms.

### 8.6 Strengthen Institutional Foundations

The economic benefits of AI depend critically on institutional quality. Property rights, contract enforcement, transparent legal systems, effective regulatory frameworks, and stable political institutions all contribute to the environment within which innovation occurs.

Strong institutions facilitate investment, reduce uncertainty, encourage entrepreneurship, and support economic growth. Consequently, institutional quality may be as important as technological capability in determining the long-run impact of AI.

### 8.7 Use AI to Improve Government Performance

Governments themselves may benefit from adopting AI technologies. Potential applications include public administration, tax administration, infrastructure management, healthcare delivery, education systems, regulatory oversight, and public-service provision.

Improved government efficiency can enhance the quality of public goods while reducing administrative costs. To the extent that AI improves governmental effectiveness, its benefits may extend beyond markets and firms to the broader society.

### 8.8 Avoid the Politics of Technological Fear

Periods of technological transformation often generate anxiety and resistance in societies. Throughout history, societies have repeatedly expressed concerns regarding mechanization, industrialization, automation, computers, and digital technologies. Many of these concerns reflected genuine adjustment costs. However, attempts to suppress technological progress have generally reduced long-run prosperity.

Public policy should therefore focus on adaptation rather than resistance. The challenge is not to prevent technological change but to help individuals, firms, and institutions adapt reasonably, timely, and successfully to it.

### 8.9 A Framework for AI Policy

The analysis of this essay suggests five broad principles for AI policy:

1. Encourage innovation.
2. Invest in human capital.
3. Facilitate economic adaptation.
4. Preserve competitive markets.
5. Strengthen institutional quality.

Together, these principles seek to maximize the productive and welfare-enhancing potential of AI while minimizing its disruptive effects.

In sum, Artificial Intelligence is likely to become one of the most important economic forces of the twenty-first century. The question facing policymakers is how societies can shape the institutional environment within which that transformation occurs. Policies that encourage innovation, support adaptation, maintain competition, and strengthen institutions are likely to generate the greatest long-run benefits. In contrast, policies that either suppress technological progress or ignore its adjustment costs may reduce the prosperity that AI is capable of creating. The ultimate success of AI will depend not only on advances in algorithms and computing power but also on the quality of the economic and political institutions that guide its adoption and use.

## 9. Summary and Conclusion

**(*What conclusions should economists, policy-makers, business leaders, and societies ultimately draw about Artificial Intelligence?*)**

Artificial Intelligence has rapidly emerged as one of the most significant economic innovations of the past few centuries. Its influence extends far beyond the realm of technology, reaching into virtually every aspect of economic life, including innovation, production, labor markets, industrial organization, public finance, and social welfare. As AI capabilities continue to advance, understanding their economic implications becomes increasingly important for firms, workers, policymakers, and society as a whole.

This essay argued that Artificial Intelligence should be viewed not merely as a technological innovation but as a multifaceted economic phenomenon. AI simultaneously functions as a form of capital, a form of synthetic labor, an economic infrastructure, a general-purpose technology, and an emerging institutional mechanism for economic decision-making. This unique combination of characteristics distinguishes AI from many previous innovations and helps explain its potentially huge transformative effects on modern economies.

The essay further demonstrated that AI possesses the potential to become a powerful multi-channel engine of productivity growth. Through automation, augmentation, optimization, prediction, and innovation, AI enhances the efficiency with which labor, capital, information, and knowledge are utilized throughout the economy. By increasing productivity and accelerating technological progress, AI can contribute significantly to long-run economic growth and rising living standards.

At the same time, AI is reshaping labor markets through a complex interaction of labor substitution, labor complementarity, and creative destruction. While some occupations and tasks are likely to disappear, new industries, occupations, and economic opportunities are also likely to emerge. The ultimate employment effects of AI will depend not solely on the jobs it destroys, but also on the jobs it creates and the capacity of workers and institutions to adapt to economic change.

The essay also examined the implications of AI for market structure and market competition. AI simultaneously generates forces that encourage market concentration and forces that encourage entrepreneurial dynamism. Data advantages, economies of scale, and network effects may strengthen dominant firms, while declining access costs and expanding technological capabilities may lower barriers to entry and create new opportunities for innovation and entrepreneurship. The future competitive landscape of the AI economy will depend on the relative strength of these two opposing forces.

From a welfare perspective, AI possesses the potential to generate substantial gains for consumers, producers, workers, governments, and society as a whole. By increasing economic efficiency, lowering costs, growing the quantity of output, improving the quality of products and services, and enhancing human capabilities, AI can significantly increase total economic surplus. Moreover, the resulting expansion of economic activity may strengthen governments' fiscal capacity and support greater provision of public goods that further contribute to economic prosperity and social well-being.

A recurring theme throughout this essay has been the importance of distinguishing between short-run disruptions and long-run outcomes. AI, like previous waves of technological change, is likely to create winners and losers during periods of adjustment. However, economic history suggests that technological progress has generally increased productivity, expanded economic opportunities, improved living standards, and generated long-run gains that substantially exceeded transitional costs. Whether AI follows this historical pattern remains one of the defining economic questions of our time, but the historical experience of previous general-purpose technologies suggests that the answer is likely to be affirmative.

Viewed through the lens of Schumpeterian creative destruction, AI represents the latest and perhaps one of the most powerful waves of innovation in the continuing evolution of market economies. Just as previous technological revolutions transformed agriculture, manufacturing, transportation, communication, and information processing, AI is transforming the economics of cognition, knowledge, and decision-making. Its effects may therefore prove broader and deeper than many previous technological advances.

Artificial Intelligence should also be understood within the broader historical narrative of human progress. Throughout history, improvements in living standards have been driven largely by humanity's ability to discover more productive ways of organizing labor, capital, knowledge, and technology. By augmenting humanity's capacity to generate knowledge, solve problems, and create value, AI has the potential to become one of the most consequential drivers of prosperity and human flourishing in modern history.

At the same time, AI itself is neither inherently beneficial nor inherently harmful. Like most transformative technologies, its ultimate consequences will depend on how it is developed, governed, adopted, and integrated into economic and social institutions. The same technology that increases productivity and expands opportunity can also generate adjustment costs, market distortions, and distributional challenges if accompanied by weak institutions or misguided policies. Consequently, the long-run impact of AI should not be viewed as technologically predetermined. Rather, it will be viewed in relation to the incentives, institutions, and policy frameworks that govern its use to ensure the prevention or minimization of its adverse effects. In this sense, the future of AI is ultimately not only a technological question, but also an economic, institutional, and societal one.

In the final analysis, the future impact of Artificial Intelligence will depend not only on advances in algorithms, computing power, and data availability, but also on the institutions, incentives, and policies that shape its development and adoption. Ultimately, societies that encourage innovation, facilitate adaptation, maintain competitive markets, invest in human capital, and strengthen institutional quality are likely to capture a greater share of AI's benefits while mitigating its associated transitional and adjustment costs.

## References:

- Becker, G. S. (1964). *Human capital: A theoretical and empirical analysis, with special reference to education*. Columbia University Press.
- Griffith, E. (2026, April 2). *How A.I. enabled two brothers to build a $1.8 billion company*. *The New York Times*. https://www.nytimes.com/2026/04/02/technology/ai-billion-dollar-company-medvi.html
- Romer, P. M. (1986). Increasing returns and long-run growth. *Journal of Political Economy, 94*(5), 1002–1037. https://doi.org/10.1086/261420
- Romer, P. M. (1990). Endogenous technological change. *Journal of Political Economy, 98*(5, Part 2), S71–S102. https://doi.org/10.1086/261725
- Schultz, T. W. (1961). Investment in human capital. *The American Economic Review, 51*(1), 1–17.
- Smith, A. (1976). *An inquiry into the nature and causes of the wealth of nations* (R. H. Campbell, A. S. Skinner, & W. B. Todd, Eds.). Oxford University Press. (Original work published 1776)
- Uzawa, H. (1965). Optimum technical change in an aggregative model of economic growth. *International Economic Review, 6*(1), 18–31. https://doi.org/10.2307/2525621
- Zeytoon-Nejad, A. (2024). *The Economics of Artificial Intelligence: Micro-, Meso-, and Macro-Economics of AI*. Essays in Economics, No. AZ-LE-007-2024-07-V1.0.
- Zeytoon-Nejad, A. (2025). Backward Growth Accounting: An Economic Tool for Strategic Planning of Business Growth. *Managerial and Decision Economics*, *46*(6), 3296-3317.
- Zeytoon-Nejad, A. (2026b). *AI and Profit: Artificial Intelligence and the Bottom Line of the Firm*. Essays in Economics, No. AZ-LE-009-2026-06-V1.0.
- Zeytoon-Nejad, A. (2026a). *Two types of dynamism under the market economy: Business cycles, creative destruction, and contingency planning for business*, Short Notes in Economics, No. AZ-SN-001-2026-05-V1.0.

## Appendix A: A Mathematical Framework for Analyzing the Net Employment Effect of Artificial Intelligence

One of the central economic questions surrounding Artificial Intelligence is whether it will ultimately increase or decrease employment. Throughout history, major technological innovations have simultaneously eliminated certain jobs while creating others. Thus, the employment effects of AI cannot be evaluated solely by observing the jobs that disappear after automation. One must also account for the new jobs that are created by the new firms, new industries, and new economic opportunities that emerge as a result of the same technological change directly and indirectly.[15] Figure A1 illustrates this process.

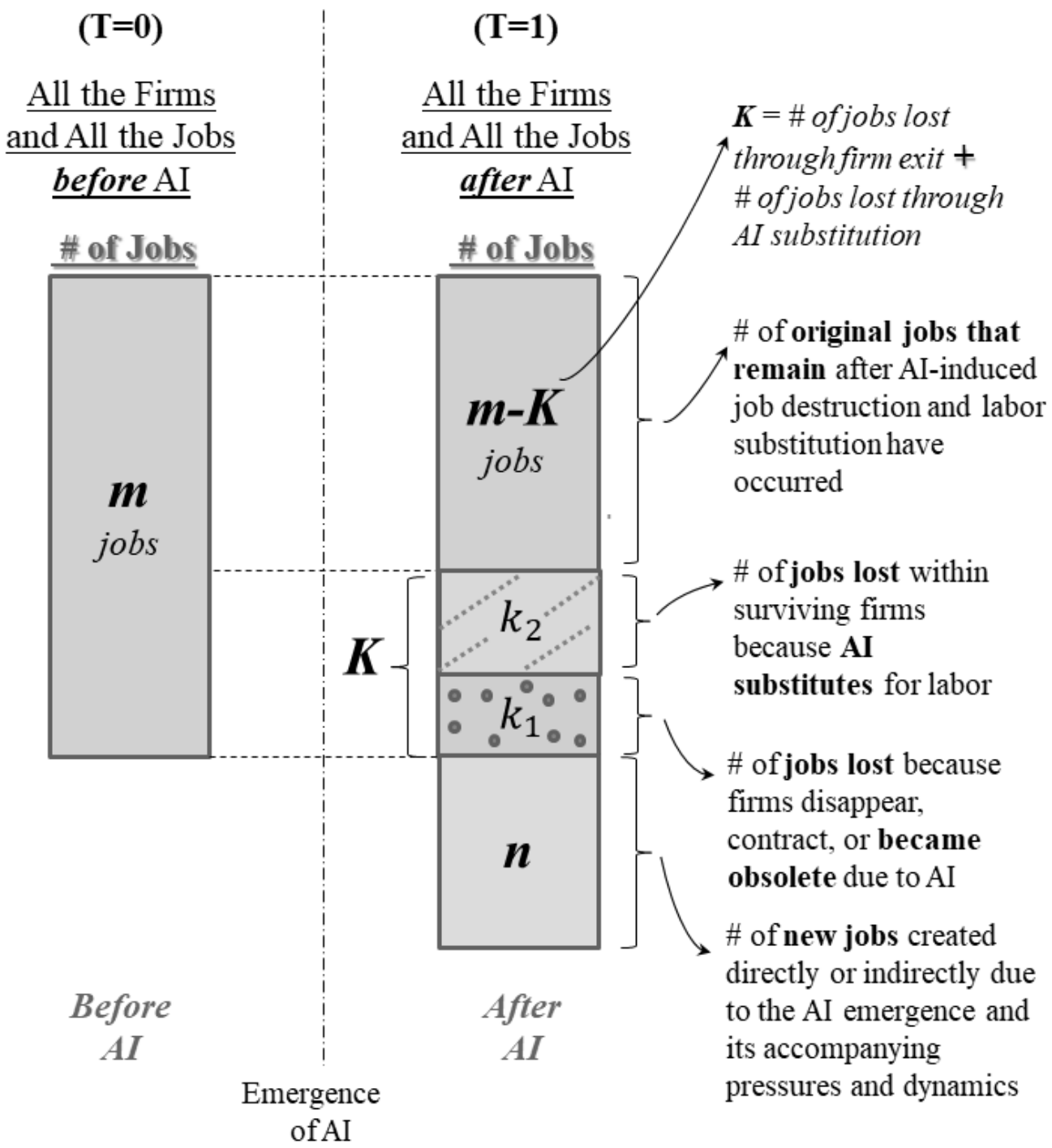

**Figure A1.** Decomposing the Employment Effects of Artificial Intelligence: Job Destruction, Labor Substitution, and New Job Creation

As shown in the above framework, prior to the emergence of AI, the economy contains a total of $m$ jobs. After the emergence of AI, some jobs disappear because firms fail ($k_1$) or because specific tasks become automated ($k_2$). At the same time, AI creates new economic opportunities that give rise to new firms and new jobs ($n$). The net employment effect depends on the relative magnitudes of these two opposing forces.

### A.1 Employment Before AI

Let total employment before the emergence of AI be denoted by:

$$E_0 = m$$

where:

- $E_0$ = total employment before AI,
- $m$ = total number of jobs in the economy prior to the emergence of AI.

This represents all jobs that existed before AI entered the economy.

### A.2 Jobs Lost After the Emergence of AI

After AI emerges, some existing jobs disappear.

Let:

$$K = k_1 + k_2$$

where:

- $k_1$ = jobs lost because firms disappear, contract, or become obsolete,
- $k_2$ = jobs lost within surviving firms because AI substitutes for labor, and
- $K$ = total number of jobs lost due to AI.

Thus:

**Total Jobs Lost =** Jobs Lost Through Firm Failure + Jobs Lost Through AI Substitution

The distinction is important. The component $k_1$ represents the classic Schumpeterian process of creative destruction in which entire firms, industries, or occupations disappear. The component $k_2$ represents labor substitution occurring within surviving firms. These firms continue operating, but some workers are replaced by AI systems, algorithms, robots, or AI agents.

[15]. Examples of directly AI-related job creation may include AI engineers, prompt engineers, data scientists specialized in AI usage, AI auditors, AI safety specialists, AI trainers, AI integration consultants, and new entrepreneurial ventures enabled by AI.

**A.3 New Jobs Created by AI**

At the same time, AI creates new opportunities.

Let n = new jobs created. These jobs may arise through new firms, new industries, new occupations, new entrepreneurial ventures, new complementary economic activities enabled by AI.

Thus:

n = New Jobs Created Through AI-Enabled Opportunities**.**

The parameter *n* captures the dynamic job-creation side of technological progress.

**A.4 Employment After AI**

After AI emerges, employment consists of the original jobs that remain, minus jobs lost through firm failure ($k_1$), minus jobs lost through labor substitution ($k_2$), plus newly created jobs ($n$).

Thus:

$$E_1 = m - k_1 - k_2 + n$$

Since

$$K = k_1 + k_2$$

the equation may be simplified to:

$$E_1 = m - K + n$$

**Employment After AI =**
Remaining Jobs from the Original Employment
**+**
New Jobs Created.

Equivalently,

$$E_1 = m + n - K$$

where $E_1$ denotes employment after AI emerges. This equation summarizes the employment structure shown in Figure A1.

**A.5 Net Employment Effect of AI**

The net employment effect is simply the difference between employment after AI and employment before AI.

Thus:

$$\Delta E_t = E_1 - E_0$$

Substituting the previous equations:

$$\Delta E_t = (m - K_1 - K_2 + n) - m$$

which simplifies to:

$$\Delta E_t = n - K_1 - K_2$$

or:

**Net Employment Change =**
New Jobs Created - Jobs Lost

or equivalently,

$$\Delta E_t = n - K$$

This equation represents the central result of the framework. The overall employment impact of AI depends entirely on whether job creation exceeds job destruction.

**A.6 Three Possible Outcomes**

**<u>Case 1:</u> Employment Increases**

If:

$$n > K$$

then:

$$\Delta E_t > 0$$

which implies:

New Jobs Created > Jobs Lost

In this case, AI creates more jobs than it destroys. Thus, total employment rises.

**<u>Case 2:</u> Employment Falls**

If:

$$n < K$$

then:

$$\Delta E_t < 0$$

which implies:

New Jobs Created < Jobs Lost

In this case, AI destroys more jobs than it creates. Thus, total employment falls.

**<u>Case 3:</u> Employment Remains Unchanged**

If:

$$n = K$$

then:

$$\Delta E_t = 0$$

which implies:

New Jobs Created = Jobs Lost

In this case, job creation exactly offsets job destruction. Hence, total employment remains unchanged.

### A.7 Historical Perspective

The historical record indicates that technological progress has often produced substantial job destruction in specific firms, occupations, and industries while simultaneously generating new jobs elsewhere in the economy. Agricultural mechanization displaced agricultural labor but helped create manufacturing employment. Industrial automation eliminated certain factory jobs but generated new opportunities in engineering, logistics, management, and services. Computers reduced demand for many clerical occupations while creating entirely new industries and professions.

From this perspective, the key economic question is not whether AI will destroy jobs—it almost certainly will. The more important question is whether the number of new jobs and opportunities created by AI will exceed the number of jobs lost through automation and creative destruction.

The framework developed in this appendix suggests that the answer depends on the relative magnitudes of $n$ and $K$. If AI generates sufficient entrepreneurship, innovation, new industries, and new forms of productive activity, then the dynamic job-creation effect may outweigh the job-destruction effect. Historical experience with previous waves of technological progress suggests that this outcome is plausible and perhaps even likely, although the magnitude and timing of these effects remain central questions in the economics of Artificial Intelligence.

## Appendix B: Winners, Adapters, and Casualties: A Framework for Understanding the Effects of Artificial Intelligence on Firms

### B.1 Introduction

Artificial Intelligence (AI) represents one of the latest and most significant waves of Schumpeterian creative destruction. Like previous transformative technologies, AI is expected to reshape industries, alter competitive dynamics, transform business models, and reallocate resources throughout the economy. However, these effects are unlikely to be uniform across firms.

Some organizations are likely to benefit directly from the expansion of AI technologies, others will adapt and transform themselves through AI adoption, some will experience substantial competitive pressure, and a subset may ultimately disappear as their traditional value propositions become obsolete. This appendix presents a conceptual framework for categorizing firms according to their likely relationship with AI and the process of creative destruction.

Importantly, the classifications presented below are illustrative rather than predictive. Firms may move between categories over time as technologies evolve, business models adapt, and competitive conditions change. The purpose of this framework is not to forecast the future of any particular company or industry but rather to illustrate the mechanisms through which AI-driven creative destruction may affect firms and industries.

### B.2 Category I: AI's Net Beneficiaries

The first category consists of firms whose products and services are directly complementary to the development, deployment, and diffusion of AI technologies. These firms stand to benefit most directly from increased AI adoption because demand for their products and services tends to rise as AI usage expands throughout the economy.

Examples include NVIDIA (AI accelerators and graphics processing units), Advanced Micro Devices (AMD) (AI processors and computing hardware), Taiwan Semiconductor Manufacturing Company (TSMC) (semiconductor fabrication), Microsoft (Azure AI infrastructure and AI services), Amazon Web Services (AWS) (cloud computing and AI infrastructure), Google Cloud (AI platforms and cloud services), Oracle Cloud Infrastructure (enterprise AI infrastructure), OpenAI (foundation models and generative AI systems), Anthropic (large language models and AI services), and Palantir (AI-enabled data analytics and decision systems).

These firms supply the computational infrastructure, semiconductors, cloud platforms, AI models, and software tools upon which much of the AI economy depends. This category is expected to experience expansion, growth, and increasing economic significance.

### B.3 Category II: AI Adopters and Business-Model Transformers

The second category consists of firms that are unlikely to be displaced by AI but are increasingly integrating AI into their operations and business models. Their competitive advantage derives not from producing AI itself but from utilizing AI to improve productivity, reduce costs, improve decision-making, and create new value for customers.

Examples include JPMorgan Chase (using AI agents in fraud detection, risk assessment, and financial analytics), Bank of America (the AI-powered virtual assistant Erica and customer service automation), Goldman Sachs (AI-assisted investment research and document analysis), Mayo Clinic (AI-supported diagnostics and treatment planning), Cleveland Clinic (AI-assisted medical imaging and clinical decision support), Walmart (inventory optimization, demand forecasting, and supply-chain management), Target (consumer demand forecasting and inventory planning), FedEx (route optimization and logistics planning), UPS (AI-assisted fleet management and delivery optimization), Boeing (predictive maintenance and manufacturing optimization), Caterpillar (equipment monitoring and predictive maintenance), and Siemens (industrial automation and smart manufacturing systems).

These companies are using AI to improve operational efficiency, decision-making, customer service, forecasting, logistics, healthcare delivery, and manufacturing processes. This category is expected to exhibit adaptation, transformation, and increased productivity.

**B.4 Category III: AI-Augmented Firms**

The third category consists of firms whose core business models may remain largely unchanged, yet whose productivity and profitability may increase substantially through AI-assisted operations. Unlike Category II firms, whose business processes may undergo significant transformation, these organizations primarily use AI as a tool for incremental improvements and operational enhancement.

Examples include Marriott International (dynamic pricing and demand forecasting), Hilton Hotels (customer service automation and revenue management), McDonald’s (order optimization and demand prediction), Starbucks (inventory planning and customer analytics), Home Depot (inventory management and customer support), Lowe’s (supply-chain optimization and demand forecasting), Delta Air Lines (flight scheduling and predictive maintenance), United Airlines (operational forecasting and logistics planning), Enterprise Holdings (fleet optimization and demand forecasting), and CBRE (real-estate analytics and market forecasting).

Although AI may improve the efficiency and profitability of these firms, it is unlikely to fundamentally alter the nature of the products and services they provide. This category is expected to experience productivity enhancement and improved competitiveness compared to their competitors that choose to not incorporate AI agents into their whole business models as an augmentation.

**B.5 Category IV: AI-Disrupted Firms**

The fourth category consists of firms whose traditional value proposition increasingly overlaps with capabilities now provided directly by AI systems. These firms may survive and even prosper if they successfully adapt, but they are likely to face significant competitive pressure and may need to redesign their business models.

Examples include Grammarly (writing assistance and grammar correction), JustAnswer (expert advice and information services), Stack Overflow (technical question-and-answer services), Chegg (educational assistance and tutoring support), Quizlet (study assistance and educational content), Rev (transcription services), English-editing companies, traditional translation agencies, copywriting agencies, content-generation firms, and routine market-research providers.[16]

Historically, these organizations generated value by providing information, writing assistance, educational support, transcription, translation, or expert consultation. Generative AI increasingly performs many of these functions at substantially lower cost, greater speed, and more convenience. These firms may remain viable, but many are likely to evolve toward higher-value, specialized, or AI-integrated offerings. As a result, this category is expected to experience business-model restructuring and strategic repositioning.

**B.6 Category V: Potential AI Casualties**

The final category consists of firms whose primary value proposition may become largely obsolete if AI systems continue improving. These organizations often compete directly against capabilities that AI can replicate quickly, cheaply, and at scale.

[16]. By contrast, platforms such as Reddit present a more nuanced case. Although AI may reduce direct traffic by answering users’ questions without requiring them to visit online forums, AI systems also rely heavily on large repositories of human-generated content such as Reddit for training and information retrieval. Therefore, AI may simultaneously compete with and increase the strategic value of such platforms.

Examples may include commodity content mills, low-value SEO-content farms, routine transcription services, basic resume-writing services, certain homework-assistance businesses, commodity data-entry firms, routine document-review providers, low-value information-intermediary businesses, certain forms of customer-support outsourcing, and certain forms of routine bookkeeping services.

Unlike firms in the previous category, these organizations often have limited opportunities for product differentiation because their services are highly standardized and easily replicated through AI systems. This category is likely to exit their markets, resort to consolidation, or accept to be replaced by AI-enabled alternatives.

**B.7 Strategic Responses to AI**

Firms facing AI-driven disruption generally have three strategic options.

First, they may '*adopt*' AI and integrate it into their existing operations. Second, they may '*differentiate*' themselves by providing products, services, expertise, trust, creativity, or human interaction that AI cannot easily replicate. Third, they may attempt to '*compete*' directly against AI by continuing to provide services that AI increasingly performs more efficiently.

Historical experience suggests that firms pursuing the first two strategies are more likely to survive and prosper. Firms whose business models depend primarily on competing directly against a superior and lower-cost technology often experience declining market share and may eventually exit the market.

Accordingly, one of the central lessons of AI-driven creative destruction is that adaptation matters. Just as workers who acquire AI-related skills may gain competitive advantages in labor markets, firms that successfully integrate AI into their operations are more likely to prosper than those that resist technological change.

In short, the rise of Artificial Intelligence is unlikely to affect all firms equally. Some organizations will emerge as major beneficiaries, others will transform themselves through adaptation, some will experience substantial disruption, and a few may disappear altogether. The process mirrors previous waves of technological change that reshaped industries and reallocated economic resources.

The central lesson of creative destruction is not that technological progress eliminates economic activity. Rather, technological progress reallocates resources from less productive uses to more productive ones. In the age of AI, the firms most likely to succeed will be those that learn to leverage AI as a complement to their capabilities rather than treating it solely as a competitor. In this sense, the future competitive landscape may be determined less by whether firms face AI and more by how effectively they adapt to it.